\documentclass[prd,aps,a4paper,twocolumn,nofootinbib]{revtex4-2}

\usepackage[utf8]{inputenc}
\usepackage[normalem]{ulem}

\usepackage{graphicx,psfrag}
\usepackage{mathrsfs}
\usepackage{amsmath,amsfonts,amssymb}
\usepackage{multirow}
\usepackage{comment}
\usepackage{bm}
\usepackage{epstopdf}
\usepackage{mathrsfs}
\usepackage{comment}
\usepackage{multirow}
\usepackage{hyperref}
\usepackage{diagbox} 
\usepackage{amssymb,amsthm,wasysym}
\usepackage{capt-of}
\usepackage{physics}
\usepackage[dvipsnames]{xcolor}
\usepackage{orcidlink}

\usepackage{makecell}

\definecolor{darkblue}{rgb}{0.0, 0.1, 0.7}
\hypersetup{
    colorlinks=true,
    linkcolor=darkblue,
    filecolor=darkblue,      
    urlcolor=darkblue,
    citecolor = darkblue,
    pdftitle={New approach to the calculation of extreme-mass-ratio inspirals with a spinning secondary}
    }

  \newcommand{\Tt}{\Tilde{t}}
\renewcommand{\Tr}{\Tilde{r}}
  \newcommand{\Tz}{\Tilde{z}}
  \newcommand{\Tphi}{\Tilde{\phi}}
  \newcommand{\Tl}{\Tilde{\lambda}}

\newcommand{\bC}{\boldsymbol{C}}
\newcommand{\bTC}{\boldsymbol{\Tilde{C}}}
\newcommand{\bOm}{\boldsymbol{\Omega}}
\newcommand{\bJ}{\boldsymbol{J}}
\newcommand{\bp}{\boldsymbol{p}}

\DeclareMathOperator{\sgn}{sgn}
\DeclareMathOperator{\elK}{\mathsf{K}}
\DeclareMathOperator{\elE}{\mathsf{E}}
\DeclareMathOperator{\elD}{\mathsf{D}}
\DeclareMathOperator{\elPi}{\mathsf{\Pi}}

\begin{document}

\title{A New Approach to the Calculation of Extreme-Mass-Ratio Inspirals with a Spinning Secondary}

\author{Viktor Skoup\'y \orcidlink{0000-0001-7475-5324}}
\email{viktor.skoupy@matfyz.cuni.cz}

\affiliation{Institute of Theoretical Physics, Faculty of Mathematics and Physics, Charles University, CZ-180 00 Prague, Czech Republic}

\begin{abstract}

Extreme-mass-ratio inspirals (EMRIs) are among the most promising sources for future space-based gravitational-wave (GW) detectors, such as the Laser Interferometer Space Antenna. To fully leverage the scientific potential, the GW templates required for parameter estimation must be modeled with high accuracy for eccentric precessing binary systems with nonzero spins. This work introduces a practical and efficient framework for incorporating the effects of secondary spin in fully generic, eccentric, and off-equatorial EMRIs to the first postadiabatic order. We utilize recently found analytic solutions for the trajectories of spinning bodies in Kerr spacetime to significantly simplify the calculation of the corresponding asymptotic GW fluxes. Furthermore, thanks to the recently proven flux-balance laws, we show how to express the rates of change of the constants of motion, including the Carter-Rüdiger constant, using asymptotic Teukolsky amplitudes and purely geodesic functions that are already established in the literature. Finally, we show how this framework performs in the case of nearly spherical inspirals and numerically demonstrate that the linear-in-spin part of the phases is gauge independent. A \textit{Wolfram Mathematica} implementation of the code developed in this work is publicly available in the \texttt{KerrSpinningFluxes} package.

\end{abstract}

\maketitle

\tableofcontents

\section{Introduction}\label{sec:intro}

Future space-based gravitational-wave (GW) observatories, most notably the Laser Interferometer Space Antenna (LISA) \cite{LISA:2017pwj,LISA:2024hlh}, TianQin \cite{TianQin:2015yph}, and Taiji \cite{Ruan:2020smc}, will open the millihertz frequency band to precision GW astronomy. In this regime, LISA will observe long-lived signals from compact binaries with large mass asymmetries, providing access to astrophysical systems and strong-field gravitational dynamics that are inaccessible to current ground-based detectors. Among the most scientifically promising sources in the LISA band are extreme-mass-ratio inspirals (EMRIs), in which a stellar-mass compact object slowly inspirals into a massive Kerr black hole residing in a galactic nucleus \cite{Barack:2006pq,Gair:2012nm, Babak:2017tow}.

EMRIs are characterized by mass ratios in the range $\epsilon = \mu/M \sim 10^{-7}$ -- $10^{-4}$, which implies that the smaller body with mass $\mu$ completes $10^{4}$ -- $10^6$ orbital cycles in the strong-field region before plunging into the massive black hole with mass $M$. The resulting GW signals encode detailed information about the spacetime geometry of the central object and offer an unparalleled opportunity to test general relativity in the highly relativistic regime, as well as to measure the mass and spin of massive black holes with extraordinary precision. However, realizing this scientific potential requires waveform models whose phase accuracy is maintained throughout the entire inspiral, often spanning several years of detector observation time.

The theoretical description of EMRIs is most naturally formulated within black hole perturbation theory \cite{Poisson:2011nh, Barack:2018yvs, Pound:2021}. In this framework, the smaller object is modeled as a point particle moving in the background spacetime of the massive black hole, which is typically described by the Kerr metric, while sourcing metric perturbations. The interaction between the particle and the metric perturbation gives rise to the gravitational self-force, which drives the slow inspiral through the loss of energy and angular momentum to gravitational radiation. Due to the extreme mass ratio, the inspiral dynamics admits a systematic expansion in the powers of $\epsilon$, enabling controlled approximations to the orbital evolution and the emitted waveform.

A particularly powerful approach to EMRI modeling exploits the separation of timescales between fast orbital motion and slow radiation-reaction-driven inspiral \cite{Hinderer:2008dm}. In the multiscale expansion, the quantities that describe the orbit are divided into slowly evolving orbital parameters $\pi_i$ and rapidly evolving phases $\Phi_j$. After averaging out the orbital-timescale oscillations using a near-identity transformation \cite{vandeMeent:2018,Lynch:2022,Lynch:2024}, the evolution equations take the form \cite{Mathews:2025}
\begin{subequations}\label{eq:multiscale_evolution}
\begin{align}
    \dv{\pi_i}{t} &= \epsilon F_i^{(0)}(\pi_k) + \epsilon^2 F_i^{(1)}(\pi_k) + \order{\epsilon^3} \,, \\
    \dv{\Phi_j}{t} &= \Omega_j^{(0)}(\pi_k) + \epsilon \Omega_j^{(1)}(\pi_k) + \order{\epsilon^2} \,,
\end{align}
\end{subequations}
where $F_i^{(n)}$ are the forcing functions and $\Omega_j^{(n)}$ are the orbital frequencies along with their subleading corrections. Due to this expansion, the orbital phases, which are closely related to the GW phases, can be expressed as
\begin{equation}
    \Phi_j(t) = \frac{1}{\epsilon} \qty( \Phi^{(0)}_j(\epsilon t) + \epsilon \Phi^{(1)}_j(\epsilon t) + \order{\epsilon^2} ) \,,
\end{equation}
where $\Phi^{(0)}_j$ and $\Phi^{(1)}_j$ are the adiabatic (0PA) and first postadiabatic (1PA) terms, respectively\footnote{Here we neglect the evolution of the primary mass and spin at 1PA order and orbital resonances at 0.5PA order, since these effects can be treated without the secondary spin effects.}. The 1PA term must be included for accurate parameter estimation, as neglecting it introduces significant biases \cite{Burke:2023lno}.

It was shown that the leading 0PA forcing functions $F^{(0)}_i$ are given by the averaged dissipative part of the first-order self-force, while the 1PA terms $F^{(1)}_i$ and $\Omega^{(1)}_j$ require the calculation of other parts of the first- and second-order self-force \cite{Hinderer:2008dm} and, importantly for this work, the secondary spin effects. Thus, we can split the 1PA terms as
\begin{subequations}\label{eq:1PA_functions}
\begin{align}
    F_i^{(1)}(\pi_k) &= \chi_\parallel F_i^{(1),\chi}(\pi_k) + F_i^{(1),\text{sf}}(\pi_k) \,, \\
    \Omega_j^{(1)}(\pi_k) &= \chi_\parallel \Omega_j^{(1),\chi}(\pi_k) + \Omega_j^{(1),\text{sf}}(\pi_k) \,,
\end{align}
\end{subequations}
where $\chi_\parallel$ is a component of the dimensionless secondary spin parallel to the orbital angular momentum. The superscript $\chi$ denotes the secondary spin effects, while the superscript ``sf'' denotes other self-force effects. This framework can be efficiently implemented for fast calculation of waveforms required in LISA data analysis in packages like FastEMRIWaveforms (\texttt{FEW}) \cite{FEW,Chua:2020stf,Katz:2021,Speri:2023,Chapman-Bird:2025}. 

An important result that simplifies the calculations is the proof of flux-balance laws. They relate the 0PA forcing term and its secondary spin correction to the GW fluxes from orbits of test spinning particles \cite{Sago:2006,Akcay:2019bvk,Mathews:2021rod,Grant:2024ivt,Mathews:2025}. This allows us to avoid the expensive calculation of metric reconstruction and regularization at the particle position and focus on calculating the asymptotic amplitudes using, e.g., the Teukolsky formalism in the frequency domain \cite{Teukolsky:1973,Drasco:2006}.

Inspirals with spinning secondaries have been the subject of sustained investigation over the past decade. Fluxes of energy and angular momentum have been calculated for circular and eccentric equatorial orbits of spinning particles in the Schwarzschild and Kerr spacetimes \cite{Tanaka:1996ht,Han:2010tp,Harms:2015ixa,Harms:2016ctx,Lukes-Gerakopoulos:2017vkj,Nagar:2019,Akcay:2019bvk,Piovano:2020zin,Mathews:2021rod,Skoupy:2021}. Building on these calculations, several works have calculated inspirals and the corresponding spin-induced phase shifts \cite{Warburton:2017,Piovano:2021,Skoupy:2022}, confirming significant dephasing. As a result, the consistent inclusion of secondary spin effects is essential for accurate EMRI waveform modeling.

Astrophysical considerations further motivate the need for waveform models capable of describing generic orbital configurations. EMRIs formed through dynamical capture or multibody interactions are expected to exhibit a broad distribution of eccentricities and spin orientations when entering the LISA band, and misalignment between the orbital angular momentum and the spin of the massive black hole generically leads to orbital-plane precession. Consequently, LISA is expected to observe signals from eccentric, inclined, and precessing inspirals, for which simplified orbital models are insufficient \cite{Mapelli:2021gyv}. Accurate modeling of such systems requires fully generic orbits in Kerr spacetime, which can be utilized in the frequency domain Teukolsky equation solvers.

This was first achieved by \citet{Drummond:2022xej,Drummond:2022efc} where the linear-in-spin part of the trajectory was calculated numerically in the frequency domain. This was later used in Ref. \cite{Skoupy:2023} to calculate the GW fluxes. Next, \citet{Piovano:2024} employed the Hamilton-Jacobi formalism developed in Refs. \cite{Witzany:2018ahb,Witzany:2019} to semianalytically solve the generic orbital motion and couple it to Teukolsky fluxes solver. The results from both approaches were then used to calculate inspirals with various levels of approximation \cite{Drummond:2024,Skoupy:2025b,Drummond:2026}.

Recently, \citet{Skoupy:2025} found an analytical solution to the trajectory of spinning particles in Kerr spacetime through a linear transformation of the worldline. Consequently, the trajectory can be expressed using quantities related to a geodesic solution with shifted constants of motion. In this work, we show how to exploit this solution in the calculation of the fluxes and inspirals. First, we simplify the calculation of the fluxes using Teukolsky formalism. The next part expands on the recent result for the flux-balance laws for the third constant of motion, the Carter-Rüdiger constant\footnote{Technically, there is no flux of the Carter constant to infinity and to the horizon, but the flux-balance laws relate the average rate of change to the asymptotic amplitudes and trajectory functionals.} \cite{Grant:2024ivt}, and the analytical solutions for the actions \cite{Witzany:2024ttz}, which are needed for the fluxes \cite{Mathews:2025}. We demonstrate how to express the fluxes using already known functions for the geodesic solution. We formulate a framework for generic inspirals of spinning particles and test it on nearly spherical inspirals. All the codes developed in this paper are publicly available in the \texttt{KerrSpinningFluxes} package \cite{KerrSpinningFluxes}.

This paper is organized as follows. In Sec. \ref{sec:trajectories} we review the solution for the motion of nonspinning and spinning particles in the Kerr spacetime and introduce a new spin gauge based on the analytical solution. Next, in Sec. \ref{sec:fluxes} we show how to simplify the calculation of the Teukolsky fluxes using the analytical solution and linearize it in spin. In Sec.~\ref{sec:inspirals} we derive the explicit form of the evolution equations for the inspiral in terms of the fluxes from the previous section and demonstrate it on the calculation of nearly spherical inspirals.  Intermediate results, such as Jacobians between various geodesic parametrizations, are presented in the Appendixes.

\subsection*{Notation}

This paper uses the mostly plus metric signature $(-,+,+,+)$ and geometrized units in which $G=c=1$. Indices of four-vectors are denoted by Greek letters, whereas indices of three-vectors or three-dimensional parameter spaces are represented by Latin letters. The Riemann tensor is defined as $R^\mu{}_{\nu\kappa\lambda} = \Gamma^\mu{}_{\nu\lambda,\kappa} - \Gamma^\mu{}_{\nu\kappa,\lambda} + \Gamma^{\mu}{}_{\rho\kappa} \Gamma^{\rho}{}_{\nu\lambda} - \Gamma^{\mu}{}_{\rho\lambda} \Gamma^{\rho}{}_{\nu\kappa}$, where the partial derivative is denoted by a comma, while the covariant derivative is denoted by a semicolon.

\section{Trajectories of spinning particles}\label{sec:trajectories}

In Mathisson's gravitational skeleton method, a small object moving in the background spacetime is represented by its multipoles. In particular, the motion of spinning bodies is described by the Mathisson-Papapetrou-Dixon (MPD) equations \cite{Mathisson:2010,Papapetrou:1951pa,Dixon:1970I}
\begin{align}
    \frac{D P^\mu}{\dd \tau} &= - \frac{1}{2} R^{\mu}{}_{\nu\kappa\lambda} \dv{x^\mu}{\tau} S^{\mu\nu} + \ldots \,, \\
    \frac{D S^{\mu\nu}}{\dd \tau} &= P^\mu \dv{x^\nu}{\tau} - \dv{x^\mu}{\tau} P^\nu + \ldots \,,
\end{align}
where $P^\mu$ is the four-momentum, $S^{\mu\nu} = S^{[\mu\nu]}$ is the spin tensor, $R^{\mu}{}_{\nu\kappa\lambda}$ is the Riemann tensor, and $\tau$ is the proper time. The equations presented here include the pole and dipole contributions. Quadrupole and higher contributions can be neglected in the context of EMRIs.

Due to the freedom in choosing the representative worldline, a spin-supplementary condition must be imposed. In this work, we use the Tulczyjew-Dixon condition $S^{\mu\nu} P_\nu = 0$ \cite{tulczyjew1959motion,Dixon:1970I}. With this condition, the mass of the particle and the magnitude of its spin are defined as 
\begin{align}
    \mu &= \sqrt{- P^\mu P_\mu} \,, & S &= \sqrt{S^{\mu\nu} S_{\mu\nu}/2} \,,
\end{align}
respectively. They are conserved throughout evolution with the MPD equations. We can define the specific four-momentum, specific spin tensor, and specific spin magnitude as
\begin{align}
    u^\mu & \equiv \frac{P^\mu}{\mu} \,, & s^{\mu\nu} & \equiv \frac{S^{\mu\nu}}{\mu} \,, & s & \equiv \frac{S}{\mu}  \,.
\end{align}

It is convenient to define a specific spin vector
\begin{align}
    s^{\mu} &= - \epsilon^{\mu\nu\kappa\lambda} u_\nu s_{\kappa\lambda}/2 \,, & s^{\mu\nu} &= \epsilon^{\mu\nu\kappa\lambda} u_\kappa s_\lambda \,.
\end{align}

It has been shown that, in the context of EMRIs, the secondary spin contribution can be truncated at linear order. This allows us to neglect quadrupole and higher contributions in the dynamics since they are $\order{s^2}$. Then, the evolution equations read
\begin{align}
    \frac{D^2 x^\mu}{\dd \tau^2} &= - \frac{1}{2} R^{\mu}{}_{\nu\kappa\lambda} \dv{x^\mu}{\tau} s^{\mu\nu} \,, & \frac{D s^{\mu}}{\dd \tau} &= 0 \,.
\end{align}

The Kerr metric in Boyer-Lindquist-like coordinates $(t,r,z=\cos(\theta),\phi)$ is
\begin{multline}
    \dd s^2 = - \frac{\Delta}{\Sigma} \qty( \dd t - a (1-z^2) \dd \phi )^2 + \frac{\Sigma}{\Delta} \dd r^2 + \frac{\Sigma}{1-z^2} \dd z^2 \\ + \frac{1-z^2}{\Sigma} \qty( a \dd t - \qty(r^2+a^2) \dd \phi )^2 \,,
\end{multline}
where $\Delta = r^2 - 2 M r + a^2$, $\Sigma = r^2 + a^2 z^2$, $M$ is the mass, $a$ is the specific angular momentum (Kerr parameter), and the horizon is located at $r_+ = M + \sqrt{M^2-a^2}$. This spacetime admits not only two Killing vectors 
\begin{equation}
    \xi_{(t)}^\mu \partial_\mu = \partial_t \,, \qquad \xi_{(\phi)}^\mu \partial_\mu = \partial_\phi \,,
\end{equation}
but also a Killing-Yano tensor
\begin{multline}
    Y_{\mu\nu} \dd x^\mu \wedge \dd x^\nu = a z \dd r \wedge \qty( \dd t - a (1-z^2) \dd \phi ) \\ + r \dd z \wedge \qty( a \dd t - \qty(r^2+a^2) \dd \phi ) \,.
\end{multline}
Thanks to these symmetries, there exist additional constants of motion \cite{Dixon:1970I,Rudiger:1981,Rudiger:1983}
\begin{subequations}\label{eq:com_definitions}
\begin{align}
    E &= - \xi^{(t)}_\mu u^\mu + \frac{1}{2} \xi^{(t)}_{\mu;\nu} s^{\mu\nu} \, , \\
    L_z &= \xi^{(\phi)}_\mu u^\mu - \frac{1}{2} \xi^{(\phi)}_{\mu;\nu} s^{\mu\nu} \, , \\
    K &= K_{\mu\nu} u^\mu u^\nu + 4 u^\mu s^{\rho\sigma} Y^\kappa{}_{\left[\mu\right.} Y_{\left.\sigma\right]\rho;\kappa}  \, , \label{eq:Carter} \\ 
    s_\parallel &= \frac{Y_{\mu\nu} u^\mu s^\nu}{\sqrt{K_{\mu\nu} u^\mu u^\nu}} \, .
\end{align}
\end{subequations}
where $K_{\mu\nu} = Y_{\mu\alpha} Y_{\nu}{}^{\alpha}$. These are usually interpreted as the specific energy with respect to infinity, the component of the specific angular momentum that is parallel to the symmetry axis, the Carter-like constant, and the component of the spin that is parallel to the orbital angular momentum. We define the orthogonal component of spin as
\begin{equation}
    s_\perp \equiv \sqrt{s^2 - s_\parallel^2} \,.
\end{equation}

\subsection{Geodesic orbits}
Let us first introduce the notation used in the description of geodesic trajectories. Thanks to the existence of the constants of motion, the equations of motion can be expressed in a separated first-order form as \cite{Carter:1968}
\begin{subequations}\label{eq:geodesic}
\begin{align}
    \dv{t}{\lambda} &= T_r^{(E,L_z)}(r) + T_z^E(z) + a L_z \, , 
    \\
    \dv{\phi}{\lambda} &= \Phi_r^{(E,L_z)}(r) + \Phi_z^{L_z}(z) - a E \, ,
    \\
    \qty( \dv{r}{\lambda} )^2 &= R^{(E,L_z,K)}(r) \, , \label{eq:ur_geo}\\
    \qty( \dv{z}{\lambda} )^2 &= Z^{(E,L_z,K)}(z) \, , \label{eq:uz_geo}\\
    \dv{\tau}{\lambda} &= \Sigma \,,
\end{align}
\end{subequations}
where $\lambda$ represents the Mino time and
\begin{align}
    T_r^{(E,L_z)}(r) & \equiv \frac{r^2 + a^2}{\Delta} P_r(r) \,, \\
    T_z^E(z) &\equiv -a^2 E (1 - z^2)\,, \\
    \Phi_r^{(E,L_z)}(r) & \equiv \frac{a}{\Delta} P_r(r)\,, \\
    \Phi_z^{L_z}(z) &\equiv \frac{L_z}{1 - z^2}\,, \\
    R^{(E,L_z,K)}(r) & \equiv P_r(r)^2 - \Delta (K + r^2)\,, \\
    Z^{(E,L_z,K)}(z) &\equiv Q - z^2 \qty( Q+L_z^2 + a^2(1-E^2)(1-z^2) ) \nonumber\\ 
    &= (1 - z^2)(K -a^2 z^2) - P_z(z)^2 \,, 
\end{align}
\begin{align}
    P_r(r) &= (r^2+a^2) E - a L_z \,, \\
    P_z(z) &= L_z - a (1-z^2) E \,.
\end{align}
with $Q = K - (L_z - a E)^2$.

In the EMRI calculations, bound orbits are often parametrized by the semi-latus rectum $p$, eccentricity $e$, and inclination parameter $x$, which are derived from the radial turning points $r_1 > r_2$ and the polar turning point $z_1$ as
\begin{align}
    r_1 &= \frac{M p}{1 - e} \,, \\
    r_2 &= \frac{M p}{1 + e} \,, \\
    x &= \sgn(L_z) \sqrt{1 - z_1^2} \,.
\end{align}
The radial and polar velocities from Eqs.~\eqref{eq:ur_geo} and \eqref{eq:uz_geo} must vanish at the turning points, i.e.,
\begin{subequations}\label{eq:conditions_turning_points}
\begin{align}
    \qty( \eval{\dv{r}{\lambda}}_{r_1} )^2 &= R^{(E,L_z,K)}(r_1) = 0 \, , \\
    \qty( \eval{\dv{r}{\lambda}}_{r_2} )^2 &= R^{(E,L_z,K)}(r_2) = 0 \, , \\
    \qty( \eval{\dv{z}{\lambda}}_{z_1} )^2 &= Z^{(E,L_z,K)}(z_1) = 0 \, . 
\end{align}
\end{subequations}
From these equations, the constants of motion can be expressed as functions of the orbital parameters $p$, $e$, and $x$.

Furthermore, the motion is periodic in Mino time with the frequencies $\Upsilon_r$, $\Upsilon_z$, and $\Upsilon_\phi$. The average rates of change of the coordinate time and proper time are $\Upsilon_t$ and $\Upsilon_\tau$, respectively. The radial and polar frequencies are given by 
\begin{subequations}\label{eq:Mino_frequencies_geo}
\begin{align}
    \Upsilon_r &= \frac{\pi}{\int_{r_2}^{r_1} R^{-1/2}(r) \dd r} \,, \\
    \Upsilon_z &= \frac{\pi}{2 \int_{0}^{z_1} Z^{-1/2}(z) \dd z} \,,
\end{align}
while the rest of the frequencies can be calculated as 
\begin{align}
    \Upsilon_{t} &= \Upsilon_{tr} + \Upsilon_{tz} \,, \\
    \Upsilon_{\phi} &= \Upsilon_{\phi r} + \Upsilon_{\phi z} \,, \\
    \Upsilon_{\tau} &= \Upsilon_{\tau r} + \Upsilon_{\tau z} \,,
\end{align}
where
\begin{align}
    \Upsilon_{tr} &= \expval{T_r(r)} \,, & \Upsilon_{tz} &= \expval{T_z(z)} + a L_z \,, \\
    \Upsilon_{\phi r} &= \expval{\Phi_r(r)} \,, & \Upsilon_{\phi z} &= \expval{\Phi_z(z)} - a E \,, \\
    \Upsilon_{\tau r} &= \expval{r^2} \,, & \Upsilon_{\tau z} &= a^2 \expval{z^2} \,.
\end{align}
\end{subequations}
Here, the angle brackets denote averaging over the radial or polar period. The coordinate-time frequencies are given by
\begin{align}
    \Omega_k = \frac{\Upsilon_k}{\Upsilon_t} \,.
\end{align}

The evolution of the coordinate and proper time and the azimuthal coordinate can be expressed as a sum of a secularly growing phase and oscillatory components as 
\begin{align}
    t(q^t, q^r, q^z) &= q^t + \Delta t_r(q^r) + \Delta t_z(q^z) \,, \\
    \phi(q^t, q^r, q^z) &= q^\phi + \Delta \phi_r(q^r) + \Delta \phi_z(q^z) \,, \\
    \tau(q^t, q^r, q^z) &= q^\tau + \Delta \tau_r(q^r) + \Delta \tau_z(q^z) \,.
\end{align}
where $q_k = \Upsilon_k \lambda$ are the homogeneous phases.

The solution of the equations of motion can be written in closed form in terms of Jacobi elliptic functions (see Refs.~\cite{Fujita:2009bp,vandeMeent:2020}).

\subsection{Solution for the spin vector}

Since the spin vector $s^\mu$ follows parallel transport along the trajectory and we can neglect the quadratic-in-spin and higher terms, we can describe the evolution of $s^\mu$ using parallel transport along a reference geodesic. The spin vector can be expressed as
\begin{equation}
    s^\mu = s_\parallel e_3^{\mu} + s_\perp \qty( e_1^\mu \cos(\psi) + e_2^\mu \sin(\psi) ) \,,
\end{equation}
where $e_A^\mu$ are the components of the Marck tetrad \cite{Marck:1983,Witzany:2019}. Here, we employ the convention of \cite{Skoupy:2025} (see Eqs. (8) therein). The precession phase $\psi(\lambda)$ then evolves via the equation
\begin{equation}
    \dv{\psi}{\lambda} = - \Psi_r(r) - \Psi_z(z) \,,
\end{equation}
where
\begin{align}
    \Psi_r(r) &\equiv \frac{\sqrt{K} P_r(r)}{K + r^2} \,, &  \Psi_z(z) &\equiv \frac{a \sqrt{K} P_z(z)}{K - a^2 z^2} \,.
\end{align}
The precession frequency with respect to Mino time is
\begin{align}
    \Upsilon_\psi = \Upsilon_{\psi r} + \Upsilon_{\psi z} = - \expval{\Psi_r(r)} - \expval{\Psi_z(z)} \,.
\end{align}
Similar to the other frequencies, the coordinate-time precession frequency is $\Omega_\psi = \Upsilon_\psi/\Upsilon_t$.

It has been shown that, to linear order in spin, quantities such as the constants of motion or frequencies are independent of the orthogonal component $s_\perp$ \cite{Witzany:2019,Drummond:2022xej,Drummond:2022efc}.

\subsection{Spinning particle orbits}

Recently, it has been shown \cite{Skoupy:2025} that by performing a linear worldline shift
\begin{align}
    \Tilde{x}^\mu &= x^\mu + \delta x^\mu , \label{eq:worldline_shift} \\
    \delta x^\mu &= s_\parallel \delta x_3^\mu + s_\perp \qty( \delta x_1^\mu \cos(\psi) + \delta x_2^\mu \sin(\psi) ) \, ,
\end{align}
the equations of motion for the new ``virtual'' worldline $\Tilde{x}^\mu(\tau)$ are separable. Here, $\delta x_A^\mu$ represents the components of a tetrad introduced in Eqs. (14) of \cite{Skoupy:2025}. After transforming the equations for the constants of motion \eqref{eq:com_definitions} to the new worldline and defining the auxiliary vector
\begin{equation}
    \Tilde{w}^\alpha = \Tilde{v}^\alpha \qty( 1 - s_\parallel \frac{3E}{2\sqrt{K}} ) + s_\parallel \frac{3 E}{2\sqrt{K}} \Tilde{\xi}^\alpha_{(t)}
\end{equation}
where $\Tilde{v}^\alpha$ is the four-velocity with respect to the virtual worldline, it can be shown that $\Tilde{w}^\alpha$ corresponds to a four-velocity of a geodesic with shifted constants of motion
\begin{align}
    \Tilde{E} &= E + \frac{s_\parallel(1 - E^2)}{2\sqrt{K}}  \, ,  \\
    \Tilde{L}_z &= L_z + \frac{s_\parallel(a - L_z E/2)}{\sqrt{K}} \, ,  \\
    \Tilde{K} &= K + \frac{s_\parallel( 3 a (L_z-aE) - K E )}{\sqrt{K}} \,.
\end{align}

The equations of motion for $\dv*{\Tilde{x}^\mu}{\tau} = \Tilde{v}^\mu$ can be expressed using the functions that appear in the geodesic equations \eqref{eq:geodesic} as
\begin{align}
    \dv{\Tt}{\Tl} &= T_r^{(\Tilde{E},\Tilde{L}_z)}(\Tr) + T_z^{\Tilde{E}}(\Tz) + a \Tilde{L}_z - \frac{3 s_\parallel \Tilde{\Sigma}}{2\sqrt{K}} \, , \label{eq:eom_t}\\
    \qty( \dv{\Tr}{\Tl} )^2 &= R^{(\Tilde{E}, \Tilde{L}_z, \Tilde{K})}(\Tr) \, , \\
    \qty( \dv{\Tz}{\Tl} )^2 &= Z^{(\Tilde{E}, \Tilde{L}_z, \Tilde{K})}(\Tz) \, , \\
    \dv{\Tphi}{\Tl} &= \Phi_r^{(\Tilde{E}, \Tilde{L}_z)}(\Tr) + \Phi_z^{\Tilde{L}_z}(\Tz) - a \Tilde{E} \, , \\
    \dv{\tau}{\Tl} &= \qty( 1 - \frac{3 s_\parallel E}{2 \sqrt{K}} ) \Tilde{\Sigma} \, .
\end{align}    
The solution parametrized by the constants $\boldsymbol{C} = (E,L_z,K)$ can thus be expressed using geodesic functions $x^\mu_{\text{g}}$, $\tau_\text{g}$ as
\begin{subequations}\label{eq:trajectory_result}
\begin{align}
    t(\Tl; \boldsymbol{C}) &= t_\text{g}(\Tl; \boldsymbol{\Tilde{C}}) - \frac{3 s_\parallel}{2 \sqrt{K}} \tau_\text{g}(\Tl; \boldsymbol{\Tilde{C}}) - \delta x^t(\Tl) \, , \\
    x^k(\Tl; \boldsymbol{C}) &= x^k_\text{g}(\Tl; \boldsymbol{\Tilde{C}}) - \delta x^k(\Tl) \, , \\
    \tau(\Tl; \boldsymbol{C}) &= \qty( 1 - \frac{3 s_\parallel E}{2 \sqrt{K}} ) \tau_\text{g}(\Tl; \boldsymbol{\Tilde{C}}) \, , \label{eq:result_tau}
\end{align}
\end{subequations}
where $\boldsymbol{\Tilde{C}} = (\Tilde{E},\Tilde{L}_z,\Tilde{K})$ are the shifted constants

Similarly, Mino-time frequencies can be expressed using geodesic expressions with shifted constants as
\begin{align}
    \Upsilon_t(\boldsymbol{C}) &= \Upsilon_t^\text{g}(\boldsymbol{\Tilde{C}}) - \frac{3 s_\parallel}{2\sqrt{K}} \Upsilon_\tau^\text{g}(\boldsymbol{\Tilde{C}}) \, , \\
    \Upsilon_k(\boldsymbol{C}) &= \Upsilon_k^\text{g}(\boldsymbol{\Tilde{C}}) \, , \\
    \Upsilon_\tau(\boldsymbol{C}) &= \qty( 1 - \frac{3 s_\parallel E}{2 \sqrt{K}} ) \Upsilon_\tau^\text{g}(\boldsymbol{\Tilde{C}}) \, .
\end{align}

\subsection{Linearization and the spin gauge}

Because we are working in the linear regime, we can separate various quantities into a geodesic part and a linear part as
\begin{equation}
    f(\boldsymbol{\pi}, s_\parallel) = f^{0,\boldsymbol{\pi}}(\boldsymbol{\pi}) + s_\parallel f^{1,\boldsymbol{\pi}}(\boldsymbol{\pi}) \,,
\end{equation}
where $\boldsymbol{\pi} = \pi_i$ are some parameters. The choice of $\pi_i$ specifies the reference geodesic. When using a different set of parameters $\boldsymbol{\varpi} = \varpi_j$, which is related to $\pi_i$ as
\begin{equation}
    \varpi_j(\boldsymbol{\pi},s_\parallel) = \varpi_j^{0,\boldsymbol{\pi}}(\boldsymbol{\pi}) + s_\parallel \varpi_j^{1,\boldsymbol{\pi}}(\boldsymbol{\pi}) \,,
\end{equation}
the linearization reads
\begin{equation}
    f(\boldsymbol{\varpi}, s_\parallel) = f^{0,\boldsymbol{\varpi}}(\boldsymbol{\varpi}) + s_\parallel f^{1,\boldsymbol{\varpi}}(\boldsymbol{\varpi}) \,.
\end{equation}
While the geodesic parts are the same, i.e.
\begin{equation}
    f^{0,\boldsymbol{\pi}}(\boldsymbol{\pi}) = f^{0,\boldsymbol{\varpi}}(\boldsymbol{\varpi}^{0,\boldsymbol{\pi}}(\boldsymbol{\pi})) \,,
\end{equation}
the linear parts are related by
\begin{equation}\label{eq:gauge_transformation}
    f^{1,\boldsymbol{\pi}}(\boldsymbol{\pi}) = f^{1,\boldsymbol{\varpi}}(\boldsymbol{\varpi}) + \pdv{f^{0,\boldsymbol{\varpi}}}{\varpi_j} \varpi_j^{1,\boldsymbol{\pi}} \,,
\end{equation}
or, alternatively,
\begin{equation}\label{eq:gauge_transformation2}
    f^{1,\boldsymbol{\pi}}(\boldsymbol{\pi}) = f^{1,\boldsymbol{\varpi}}(\boldsymbol{\varpi}) - \pdv{f^{0,\boldsymbol{\pi}}}{\pi_j} \pi_j^{1,\boldsymbol{\varpi}} 
\end{equation}
when it is easier to calculate $\dv*{f^0}{\pi_j}$ than $\dv*{f^0}{\varpi_j}$. From the equations above we can see that
\begin{equation}
    \varpi_j^{1,\boldsymbol{\pi}} = - \pdv{\varpi^0_j}{\pi_k} \pi_k^{1,\boldsymbol{\varpi}} \,.
\end{equation}
Parameters $\boldsymbol{\pi}$ and $\boldsymbol{\varpi}$ can be, for example, orbital parameters $\bp \equiv (p,e,x)$ (calculated from the average turning points in the spinning case), constants of motion $\bC \equiv (E,L_z,K)$\footnote{In Refs. \cite{Witzany:2019,Piovano:2024} a modified set of constants $(E,L_z,K - 2 a s_\parallel \sgn(L_z-a E))$ was held fixed to regularize the linear parts in the equatorial limit.}, orbital frequencies $\bOm \equiv (\Omega_r,\Omega_z,\Omega_\phi)$\footnote{Parametrization with the frequencies fails near the separatrix. See Appendix \ref{app:jacobians} for details.}, actions $\bJ \equiv (J_r, J_\theta, J_\phi)$ \cite{Witzany:2024ttz}, etc. In practice, even when a different set of parameters than $(p,e,x)$ is used, the functions $f^0$ and $f^1$ are given in terms of $(p,e,x)$ which are related to the reference geodesic.

In the calculation of EMRI waveforms, the choice of the reference geodesic is pure gauge. Ref \cite{Mathews:2025} showed that the full 1PA waveform is invariant under such reparametrizations since it only redistributes the subleading corrections between the forcing functions, the frequencies and the waveform amplitudes. This leads to nonzero differences only at $\order{\epsilon^3}$. 

Different choices of a reference geodesic are advantageous in different cases. For example, for equatorial orbits, it is convenient to use an equatorial reference geodesic and match the energy $E$ and angular momentum $L_z$ as the remaining degrees of freedom. However, if we consider generic off-equatorial orbits and use the three constants $E$, $L_z$, $K$ as our parameters, the linear parts are generally different from those in the previous choice for equatorial orbits. Furthermore, the linear parts in the generic off-equatorial case with fixed constants diverge near special orbits, such as equatorial and spherical orbits, as well as the separatrix. These divergences can be partially removed by fixing different combinations of parameters such as $K - 2 a s_\parallel \sgn(L_z-aE)$ \cite{Witzany:2019,Piovano:2024}, or by requiring that the corrections to $r_2$ and $r_3$ are identical \cite{Piovano:2025}.

\subsubsection{Fixed-virtual-worldline gauge}\label{sec:CTilde_gauge}

Here, we propose using a gauge in which we fix the shifted constants of motion $\boldsymbol{\Tilde{C}} \equiv (\Tilde{E},\Tilde{L}_z,\Tilde{K})$. Then, it is straightforward to linearize the trajectory with this choice of reference geodesics. 
The linear parts of the constants are simply
\begin{subequations}\label{eq:constants_corrections}
\begin{align}
    E^{1,\bTC} &= - \frac{1-\Tilde{E}^2}{2\sqrt{\Tilde{K}}} \,, \\
    L_z^{1,\bTC} &= - \frac{a- \Tilde{L}_z \Tilde{E}/2}{\sqrt{\Tilde{K}}} \,, \\
    K^{1,\bTC} &= - \frac{3 a (\Tilde{L}_z - a \Tilde{E}) - \Tilde{K} \Tilde{E}}{\sqrt{\Tilde{K}}} \,.
\end{align}
\end{subequations}
From Eqs.~\eqref{eq:trajectory_result}, we can see that the linear parts of the trajectory parametrized by Mino time are
\begin{align}
    t^{1,\bTC}(\Tl) &= - \frac{3}{2\sqrt{\Tilde{K}}} \tau^0(\Tl) - \delta x^t \,, \\
    r^{1,\bTC}(\Tl) &= - \delta x^r \,, \\
    z^{1,\bTC}(\Tl) &= - \delta x^z \,, \\
    \phi^{1,\bTC}(\Tl) &= - \delta x^\phi \,,
\end{align}
while the linear parts of the Mino-time frequencies read
\begin{align}
    \Upsilon_t^{1,\bTC} &= -\frac{3}{2\sqrt{\Tilde{K}}} \Upsilon_\tau^0 \,, \\
    \Upsilon_r^{1,\bTC} &= 0 \,, \\
    \Upsilon_z^{1,\bTC} &= 0 \,, \\
    \Upsilon_\phi^{1,\bTC} &= 0 \,.
\end{align}
From these expressions, the linear parts of the coordinate frequencies can be calculated as
\begin{equation}
    \Omega_k^{1,\bTC} = - \Omega_k^0 \frac{\Upsilon_t^{1,\bTC}}{\Upsilon_t^0} = \Omega_k^0 \frac{3 \Upsilon_\tau^0}{2 \Upsilon_t^0\sqrt{\Tilde{K}}} \,. \label{eq:frequencies_corrections}
\end{equation}
Note that the linear part is calculated as the geodesic part multiplied by a factor that is the same for all coordinate frequencies and is proportional to the inverse of the average redshift. 

This gauge is advantageous because there are no divergences near special orbits that appear in the fixed constants and fixed frequencies gauges. This can be easily seen from Eqs.~\eqref{eq:constants_corrections} -- \eqref{eq:frequencies_corrections} and from the fact that the reference geodesic is of the same type as the physical trajectory (e.g., eccentric, off-equatorial, marginally bound, etc.). 

Unless stated otherwise, this fixed $\bTC$ gauge will be used throughout this paper, and we will use parametrization with orbital parameters $\boldsymbol{\Tilde{p}} \equiv (\Tilde{p},\Tilde{e},\Tilde{x})$ that are geodesically related to the shifted constants $\bTC$ through the relations \eqref{eq:conditions_turning_points}, where $\boldsymbol{C}$ is replaced with $\boldsymbol{\Tilde{C}}$.

\subsubsection{Transformation to different gauges}\label{sec:gauge_transformation}

To transform the linear parts from the fixed $\bTC$ gauge to different gauges or vice versa, Eqs.~\eqref{eq:gauge_transformation} or \eqref{eq:gauge_transformation2} can be utilized. In particular, when transforming to the fixed constants gauge, the relation reads
\begin{equation}\label{eq:transformation_Ctilde}
    f^{1,\bC} = f^{1,\bTC} - \pdv{f^0}{E} E^{1,\bTC} - \pdv{f^0}{L_z} L_z^{1,\bTC} - \pdv{f^0}{K} K^{1,\bTC}
\end{equation}
where the linear parts of the constants of motion from Eq.~\eqref{eq:constants_corrections} are utilized.

Similarly, for transformation to fixed average turning points, it holds 
\begin{multline}\label{eq:transformation_ptilde}
    f^{1,\bp}(\boldsymbol{\Tilde{p}}) = f^{1,\bTC}(\boldsymbol{\Tilde{p}}) \\ - \pdv{f^0}{p} p^{1,\bTC}(\boldsymbol{\Tilde{p}}) - \pdv{f^0}{e} e^{1,\bTC}(\boldsymbol{\Tilde{p}}) - \pdv{f^0}{x} x^{1,\bTC}(\boldsymbol{\Tilde{p}}) \,,
\end{multline}
For Eq.~\eqref{eq:transformation_ptilde}, we need to find the correction to the parameters $p^1, e^1, x^1$ when they are written in terms of the virtual-worldline parameters as 
\begin{align}
    p &= \Tilde{p} + s_\parallel p^1(\Tilde{p},\Tilde{e},\Tilde{x}) \\
    e &= \Tilde{e} + s_\parallel e^1(\Tilde{p},\Tilde{e},\Tilde{x}) \\
    x &= \Tilde{x} + s_\parallel x^1(\Tilde{p},\Tilde{e},\Tilde{x})
\end{align}
This can be achieved by calculating corrections to the turning points of the physical worldline $r_1^1$, $r_2^1$, and $z_1^1$, which are related to the corrections to the orbital parameters through the Jacobian in Eqs.~\eqref{eq:Jacobian_pex_tp}.

The average turning points of the physical worldline can be expressed from Eq.~\eqref{eq:worldline_shift} as\footnote{This holds because the components $\delta x^r$ and $\delta x^z$ have maxima or minima at the turning points, and thus the virtual and physical trajectories reach the turning points simultaneously.}
\begin{align}
    \expval{r_1} &= \Tr_1 - \expval{\delta x^r(\Tr_1)}_{\Tz,\psi} \,, \\
    \expval{r_2} &= \Tr_2 - \expval{\delta x^r(\Tr_2)}_{\Tz,\psi} \,, \\
    \expval{z_1} &= \Tz_1 - \expval{\delta x^z(\Tz_1)}_{\Tr,\psi} \,.
\end{align}
where the averages of $\delta x^r$ are taken over the polar motion $\Tz(q_z)$ and the precession phase $\psi$, and the average over $\delta x^z$ is taken over the radial motion $\Tr(q_r)$ and the precession phase $\psi$. Because the contribution from the precession phase is purely oscillatory, the resulting averages are independent of $s_\perp$. The corrections to the turning points are, thus,
\begin{align}
    r_1^1 &= - \expval{\delta x_3^r(\Tr_1)}_{\Tz} \,, \\
    r_2^1 &= - \expval{\delta x_3^r(\Tr_2)}_{\Tz} \,, \\
    z_1^1 &= - \expval{\delta x_3^z(\Tz_1)}_{\Tr} \,.
\end{align}

By substituting the coordinate expressions for $\delta x_3^\mu$, these relations can be written as
\begin{subequations}\label{eq:turning_points_corrections}
\begin{align}
    r_{1,2}^1 &= - \frac{P_r(\Tr_{1,2})}{\Tr_{1,2} \sqrt{\Tilde{K}}} \expval{\frac{\Tr_{1,2}^2}{\Tr_{1,2}^2 + a^2 \Tz^2}}_{\Tz} \,, \\
    z_1^1 &= - \frac{P_z(\Tz_1)}{\sqrt{\Tilde{K}}} \expval{\frac{a \Tz_1}{\Tr^2+a^2 \Tz_1^2}}_{\Tr} \,,
\end{align}
\end{subequations}
where the averages can be explicitly expressed in terms of Legendre elliptic integrals as
\begin{align}
    \expval{\frac{\Tr_{1,2}^2}{\Tr_{1,2}^2 + a^2 \Tz^2}}_z &= \frac{\elPi\qty(\alpha_z^2(\Tr_{1,2}), k_z)}{\elK(k_z)} \,, \\
    \expval{\frac{a \Tz_1}{r^2 + a^2 \Tz_1^2}}_r &= \frac{a \Tz_1}{\Tr_3^2 + a^2 \Tz_1^2} \nonumber \\
     &\phantom{=} + \Im{ \frac{(\Tr_2 - \Tr_3) \elPi\qty( \alpha_r^2, k_r)}{2 (\Tr_2 + i a \Tz_1)(\Tr_3 + i a \Tz_1) \elK(k_r)} } \,.
\end{align}
Here $\elK(k)$ and $\elPi(\alpha^2,k)$ represent the elliptic integrals of the first and third kind, respectively, and
\begin{align}
    \alpha_z^2(\Tr) &= - \frac{a^2 \Tz_1^2}{\Tr^2} \,, \\
    k_z^2 &= \frac{\Tz_1^2}{\Tz_2^2} \,, \\
    \alpha_r^2 &= \frac{(\Tr_1 - \Tr_2)(\Tr_3 + i a \Tz_1)}{(\Tr_1 - \Tr_3)(\Tr_2 + i a \Tz_1)} \,, \\
    k_r^2 &= \frac{(\Tr_1 - \Tr_2)(\Tr_3 - \Tr_4)}{(\Tr_1 - \Tr_3)(\Tr_2 - \Tr_4)} \,.
\end{align}
The radial turning points $\Tr_i$ can be found in \cite{vandeMeent:2020} while the polar turning points read
\begin{align}
    \Tz_1 &= \sqrt{1-\Tilde{x}^2} \,, \\
    a \Tz_2 &= \sqrt{ a^2 (1 - \Tz_1^2) + \frac{\Tilde{Q}^2 + \Tilde{L}_z^2}{1-\Tilde{E}^2} } \,,
\end{align}
When the combination $a \Tz_2$ instead of $\Tz_2$ is used, the result is regular in the Schwarzschild case $a \rightarrow 0$.

\begin{table}[t!]
    \centering
    \begin{tabular}{c|ccccc}
         & $\bTC$ & $\bC$ & $\bp$ & $\bOm$ & $\bJ$ \\ \hline
        $\bTC$ &  & $\bTC^{1,\bC}$ & $\bTC^{1,\bp}$ & $\bTC^{1,\bOm}$ & $\bC^{1,\bJ}$ \\
        $\bC$ & \fbox{$\bC^{1,\bTC}$} &  & $\bC^{1,\bp}$ & $\bC^{1,\bOm}$ & $\bC^{1,\bJ}$ \\
        $\bp$ & \fbox{$\bp^{1,\bTC}$} & $\bp^{1,\bC}$ &  & $\bp^{1,\bOm}$ & $\bp^{1,\bJ}$ \\
        $\bOm$ & \fbox{$\bOm^{1,\bTC}$} & $\bOm^{1,\bC}$ &$\bOm^{1,\bp}$ &  & $\bOm^{1,\bJ}$ \\
        $\bJ$ & $\bJ^{1,\bTC}$ & \fbox{$\bJ^{1,\bC}$} &$\bJ^{1,\bp}$ & $\bJ^{1,\bOm}$ & 
    \end{tabular}
    \caption{Linear parts of different quantities (rows) in different gauges (columns). The boxed expressions are derived in Secs.~\ref{sec:CTilde_gauge} and \ref{sec:gauge_transformation} of this paper or already known analytically. From them, all other expressions can be calculated. Explicit formulas are presented in Appendix \ref{app:linear_parts}.}
    \label{tab:linear_parts}
\end{table}

It is easy to use Eqs.~\eqref{eq:gauge_transformation} and \eqref{eq:gauge_transformation2} to find the transformation to other gauges from the analytic expressions in Eqs.~\eqref{eq:frequencies_corrections}, \eqref{eq:constants_corrections}, and \eqref{eq:turning_points_corrections}, as well as from the analytical expressions for the actions $J_r(\bC,s_\parallel)$ and $J_z(\bC,s_\parallel)$ from \cite{Witzany:2024ttz}. A list of necessary quantities is provided in Table \ref{tab:linear_parts}. In the transformation relations, we also need the derivatives of the geodesic orbital parameters $p$, $e$, $x$, geodesic frequencies, and actions, with respect to the constants of motion. These can be found in Appendix \ref{app:jacobians}. A list of all expressions for the quantities in Table \ref{tab:linear_parts} is provided in Appendix \ref{app:linear_parts}. We verified the validity of our expressions by comparing the numerical results for fixed turning points\footnote{The definition of fixed turning points gauge follows Refs. \cite{Drummond:2022efc,Skoupy:2023,Piovano:2024,Witzany:2019,Skoupy:2025b} and does not correspond to the identically named gauge in \cite{Mathews:2025}.} and fixed constants gauges with the results from Refs.~\cite{Drummond:2022efc,Piovano:2024,Witzany:2024ttz}. A \textit{Wolfram Mathematica} implementation of all formulas is also available in \cite{KerrSpinningFluxes}.

\begin{figure}
    \centering
    \includegraphics[width=\linewidth]{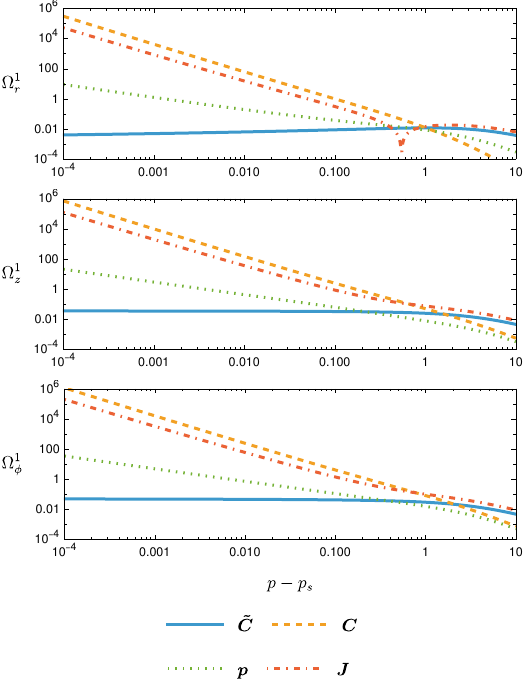}
    \caption{Linear parts of the frequencies $\Omega_i^1$ in different spin gauges: fixed shifted constants $\bTC$, fixed constants $\bC$, fixed turning points $\bp$, and fixed actions $\bJ$. The orbital parameters are $a=0.95M$, $e=0.5$, $x=0.5$.}
    \label{fig:delta_Omega}
\end{figure}

To illustrate the transformation between the gauges and the regularity of linear parts in the virtual-worldline gauge, we plot the linear parts of the frequencies in different spin gauges near the separatrix in Fig.~\ref{fig:delta_Omega}. We can see that in the fixed constants, fixed turning points, and fixed actions gauges the linear parts diverge near the separatrix, while in the fixed $\bTC$ gauge, it remains finite, and $\Omega_r^{1,\bTC}$ approaches zero at the separatrix.

\section{Gravitational-wave fluxes}\label{sec:fluxes}

In this section, we demonstrate how to exploit the analytic solution for the trajectory to significantly simplify the calculation of energy and angular momentum fluxes through the Teukolsky amplitudes. Moreover, by utilizing recently found formulas for the flux of the Carter-like constant $K$, we show how to express this flux using geodesic functions.

\subsection{Teukolsky amplitudes}

In the Newman-Penrose formalism, gravitational perturbations are calculated from the Weyl scalar
\begin{equation}
    \Psi_4 = C_{\alpha\beta\gamma\delta} \, n^\alpha \bar{m}^\beta n^\gamma \bar{m}^\delta \,,
\end{equation}
where $C_{\alpha\beta\gamma\delta}$ represents the Weyl tensor, and $n^\alpha$ along with $\bar{m}^\beta$ are components of the Kinnersley tetrad. The perturbation of the Weyl scalar $\psi_4$ around the Kerr background satisfies the Teukolsky equation
\begin{equation}\label{eq:Teukolsky_full}
    \mathcal{O}' \psi_4 = 8 \pi \mathcal{S}_4 T \,,
\end{equation}
where $\mathcal{O}'$ is a second-order partial differential operator and $\mathcal{S}_4 T$ is another differential operator that acts on the stress-energy tensor \cite{Pound:2021}. The Weyl scalar can be decomposed into frequencies $\omega$ and multipoles $l,m$ as
\begin{equation}
    \zeta^4 \psi_4 = \int \sum_{l,m} \psi_{lm\omega}(r) \, _{-2}S_{lm}^{a\omega}(z) e^{-i \omega t + i m \phi} \dd \omega \,,
\end{equation}
where $\zeta = r - i a z$, $\psi_{lm\omega}(r)$ is the radial function, and $_{-2}S_{lm}^{a\omega}(z)$ is the spin-weighted spheroidal harmonic. The radial function satisfies the radial Teukolsky equation
\begin{multline}
    \Delta^2 \dv{r} \qty( \Delta^{-1} \dv{\psi_{lm\omega}}{r} ) \\ + \qty( \frac{\mathcal{K}_r^2 + 4 i (r-M) \mathcal{K}_r}{\Delta} - 8 i \omega r - \lambda_{lm\omega} )\psi_{lm\omega} = \mathcal{T}_{lm\omega} \,,
\end{multline}
where $\mathcal{K}_r = (r^2+a^2) \omega - a m$ and $\mathcal{T}_{lm\omega}$ is the decomposition of the original source from Eq.~\eqref{eq:Teukolsky_full} into frequencies $\omega$ and spin-weighted spheroidal harmonics in the form
\begin{equation}\label{eq:source_radial}
    \mathcal{T}_{lm\omega} = - 8 \int \dd t\! \int \dd z\! \int \dd \phi \, \Sigma \zeta^4 (\mathcal{S}_4 T) {}_{-2}S_{lm}^{a\omega}(z) e^{i \omega t - i m \phi} \,.
\end{equation}

The radial function behaves asymptotically as
\begin{align}
    \psi_{lm\omega}(r) &= C^+_{lm\omega} r^3 \exp(i \omega r^\ast) & \text{for } r &\rightarrow \infty \,, \\
    \psi_{lm\omega}(r) &= C^-_{lm\omega} \Delta^2 \exp(- i k_\mathcal{H} r^\ast) & \text{for } r &\rightarrow r_+ \,,
\end{align}
where $k_\mathcal{H} = \omega - m a/(2 M r_+)$ is the frequency at the horizon and $r^\ast = \int (r^2+a^2)/\Delta \dd r$ is the tortoise coordinate. The asymptotic amplitudes $C^\pm_{lm\omega}$ can be found by the variation of constants as an integral over the source term $\mathcal{T}_{lm\omega}$ 
\begin{equation}
    C^\pm_{lm\omega} = \frac{1}{W_{lm\omega}} \int_{r_+}^\infty \frac{R^\mp_{lm\omega}(r')}{\Delta^2(r')} \mathcal{T}_{lm\omega}(r') \dd r' \,,
\end{equation}
where $R^\pm_{lm\omega}$ are the homogeneous solutions that satisfy purely outgoing boundary conditions at infinity and purely ingoing boundary conditions at the horizon, respectively, with transmission amplitudes normalized to unity, and 
\begin{equation}
    W_{lm\omega} = \frac{1}{\Delta} \qty( R^+_{lm\omega} \dv{R^-_{lm\omega}}{r} - \dv{R^+_{lm\omega}}{r} R^-_{lm\omega} ) = \text{const}
\end{equation}
is the constant Wronskian.

Because the source term $\mathcal{T}_{lm\omega}$ in Eq.~\eqref{eq:source_radial} contains integration over $t, z, \phi$, we can write the amplitude as an integral over spacetime. From integration by parts, the amplitudes can be written as
\begin{equation}\label{eq:Cpm_Phi}
    C^\pm_{lm\omega} = \frac{1}{W_{lm\omega}} \int T^{\mu\nu} \Phi^{\pm}_{\mu\nu} \sqrt{-g}\, \dd x^4 \,,
\end{equation}
where
\begin{multline}
    \Phi^\pm_{\mu\nu}(x^\alpha) = \frac{1}{\Sigma} \big( n_\mu n_{\nu} F^\pm_{nn} + n_{(\mu} \bar{m}_{\nu)} F^\pm_{n\bar{m}} \\ + \bar{m}_\mu \bar{m}_\nu F^\pm_{\bar{m}\bar{m}} \big) e^{i \omega t - i m \phi} \,.
\end{multline}
The functions $F^\pm_{ab}(r,z)$, which contain the homogeneous radial solutions and the spheroidal harmonics and their derivatives, can be found in Appendix \ref{app:source}.

The stress-energy tensor of a spinning body in the pole-dipole approximation reads
\begin{equation}
    T^{\mu\nu} = \int \frac{v^{(\mu} P^{\nu)} \delta^4}{\sqrt{-g}} \dd \tau - \nabla_\rho \int \frac{S^{\rho(\mu} u^{\nu)} \delta^4}{\sqrt{-g}} \dd \tau \,,
\end{equation}
where $\delta^4 = \delta^4(x^\alpha - x^\alpha_\text{p}(\tau))$ and $v^\mu = \dv*{x^\mu}{\tau}$. After substituting it into Eq.~\eqref{eq:Cpm_Phi} and integrating by parts, we obtain
\begin{multline}
    C^\pm_{lm\omega} = \frac{1}{W_{lm\omega}} \int \Big( v^{(\mu}(\tau) P^{\nu)}(\tau) \Phi_{\mu\nu}^\pm(x^\alpha_\text{p}(\tau)) \\ + S^{\rho(\mu}(\tau) v^{\nu)}(\tau) \Phi_{\mu\nu;\rho}^\pm(x^\alpha_\text{p}(\tau)) \Big)  \dd \tau \,.
\end{multline}
Since the stress-energy tensor is invariant under the change of the representative worldline, we can write the amplitude as
\begin{equation}\label{eq:Cpm_tilde_general}
    C^\pm_{lm\omega} = \frac{1}{W_{lm\omega}} \int \qty( \Tilde{v}^{(\alpha} P^{\beta)} \Tilde{\Phi}^\pm_{\alpha\beta} +  \Tilde{S}^{\gamma(\alpha} \Tilde{v}^{\beta)} \Tilde{\Phi}^\pm_{\alpha\beta;\gamma} )  \dd \tau \,,
\end{equation}
where the quantities with a tilde correspond to the virtual worldline. Therefore, we can use the quantities associated with the virtual worldline to calculate the amplitude. Note that the four-momentum $P^\mu$ does not change under the transformation and must instead be expressed as
\begin{align}
    P^\mu &= v^\mu = \Tilde{v}^\mu - \delta v^\mu \,, \label{eq:P_from_utilde}\\
    \delta v^\mu &= \frac{D \delta x^\mu}{\dd \tau} = - \frac{s_\parallel}{\sqrt{K}} \qty( \xi^\mu_{(t)} - E u^\mu ) + \order{s_\perp} \,.
\end{align}
The other quantities associated with the virtual worldline can be easily expressed as
\begin{align}
    \Tilde{v}^\mu &= \Tilde{w}^\mu \qty( 1 + s_\parallel \frac{3 E}{2\sqrt{K}} ) -  s_\parallel \frac{3}{2\sqrt{K}} \Tilde{\xi}^\mu_{(t)} \,, \\
    \Tilde{s}^{\mu\nu} &= -\frac{s_\parallel}{\sqrt{K}} \Tilde{f}^{\mu\nu} + \order{s_\perp} \,, \\
    \dd \tau &= \qty( 1 - s_\parallel \frac{3 E}{2\sqrt{K}} ) \Tilde{\Sigma} \dd \Tilde{\lambda} \,, \label{eq:dtau}
\end{align}
where $\Tilde{f}^{\mu\nu} = -\epsilon^{\mu\nu\kappa\lambda} \Tilde{Y}_{\kappa\lambda}/2$ is the conformal Killing-Yano tensor.

Due to the multiperiodicity of the orbit, the frequency spectrum is discrete, and the amplitudes can be decomposed into partial amplitudes \cite{Drasco:2006,Skoupy:2023}
\begin{align}
    C^\pm_{lm\omega} &= \sum_{nkj} C^\pm_{lmnkj} \delta(\omega - \omega_{mnkj}) \,, \\
    \omega_{mnkj} &= m \Omega_\phi + n \Omega_r + k \Omega_z + j \Omega_\psi \,.
\end{align}
For EMRIs in the 1PA order, we can only calculate the contribution from the parallel part $s_\parallel$, as the orthogonal part $s_\perp$ appears at higher orders. This allows us to neglect the $j=\pm 1$ modes and drop the $j$ index (see Sec. III.B. of \cite{Skoupy:2025b} for discussion). 

\begin{widetext}
When we substitute the expressions \eqref{eq:P_from_utilde} -- \eqref{eq:dtau} into Eq.~\eqref{eq:Cpm_tilde_general}, after some simplifications, we obtain
\begin{equation}\label{eq:Cpmlmnk_final}
    C^\pm_{lmnk} = \frac{1 + s_\parallel \frac{E}{2 \sqrt{K}}}{2\pi \Upsilon_t W_{lm\omega}} \int \dd q^2 \qty( \Tilde{w}_n \Tilde{w}_n F_{nn}^\pm + \Tilde{w}_n \Tilde{w}_{\bar{m}} F_{n\bar{m}}^\pm + \Tilde{w}_{\bar{m}} \Tilde{w}_{\bar{m}} F_{\bar{m}\bar{m}}^\pm + \frac{s_\parallel}{\sqrt{K}} \qty( \Tilde{w}_n G_n^\pm + \Tilde{w}_{\bar{m}} G_{\bar{m}}^\pm ) )  e^{i \varphi_r(q_r) + i \varphi_z(q_z)} \,,
\end{equation}
where $\Tilde{w}_a$ is the component of the geodesiclike four-velocity in the Kinnersley tetrad, $F^\pm_{ab} = F^\pm_{ab}(\Tr(q_r),\Tz(q_z))$ with $G_{a} = G_a(\Tr(q_r),\Tz(q_z))$ are defined in Appendix \ref{app:source}, and
\begin{align}
    \varphi_{r}(q_r) &= \omega_{mnk} \Delta \Tt_r(q_r) - m \Delta \Tphi_r(q_r) + n q_r \,, \\
    \varphi_{z}(q_z) &= \omega_{mnk} \Delta \Tt_z(q_z) - m \Delta \Tphi_z(q_z) + k q_z \,.
\end{align}

The amplitudes can be linearized in the fixed $\bTC$ gauge. The advantage of this gauge is that the spatial coordinates $\Tilde{x}^i$ and velocities $\Tilde{w}^\mu$ do not have any spin correction. The only linear corrections are those to $\Upsilon_t$ in the denominator, to $\omega_{mnk}$ in $W_{lm\omega}$ and $F^\pm_{ab}$, and to $\Delta t$ in the exponential term (the terms subleading in $s_\parallel$ do not have to be expanded). The final expression for the linear part then reads
\begin{multline}
    C^{\pm,1}_{lmnk} = \frac{1}{2\sqrt{\Tilde{K}}} \qty( \Tilde{E} + 3 \frac{\Upsilon_\tau^0}{\Upsilon_t^0} \qty( 1 - \omega_{mnk}^0 \frac{\partial_\omega W^0_{lm\omega}}{W^0_{lm\omega}} ) ) C^{\pm,0}_{lmnk} \\ 
    + \frac{1}{2\pi \sqrt{\Tilde{K}} \Upsilon_t^0 W^0_{lm\omega}} \int \dd q^2 \qty( \Tilde{w}_n^0 G_n^0 + \Tilde{w}_{\bar{m}}^0 G_{\bar{m}}^0 + \omega_{mnk}^0 \frac{3\Upsilon_\tau^0}{2\Upsilon_t^0} \sum_{ab} \Tilde{w}_a^0 \Tilde{w}_b^0 \qty( \partial_\omega F_{ab}^0 + i F_{ab}^0 \qty( \Delta t^0 - \frac{\Upsilon_t^0}{\Upsilon_\tau^0} \Delta \tau^0 ) ) ) e^{i \varphi_r^0 + i \varphi_z^0} \,.
\end{multline}
\end{widetext}

\subsection{Fluxes of the constants of motion}

The flux-balance laws for energy and angular momentum were proven in \cite{Akcay:2019bvk,Mathews:2021rod} and are given by\footnote{Here we denote the fluxes by the script $\mathcal{F}$ and the general forcing function by the regular $F$.}
\begin{subequations}\label{eq:fluxes_En_Lz}
\begin{align}
    \mathcal{F}^E &= - \dv{E}{t} = \sum_{lmnk} \omega_{mnk} Z_{lmnk} \,, \\
    \mathcal{F}^{L_z} &= -\dv{L_z}{t} = \sum_{lmnk} m Z_{lmnk} \, , 
\end{align}
\end{subequations}
where 
\begin{equation}
    Z_{lmnk} = \frac{\abs{C^+_{lmnk}}^2 + \alpha_{lmnk}\abs{C^-_{lmnk}}^2}{4 \pi \omega_{mnk}^3}
\end{equation}
is the combination of the infinity and horizon amplitudes and $\alpha_{lmnk}$ is defined in Eq.~(57) of \cite{Skoupy:2023}. Recently, the flux-balance law for the Carter constant was proven in \cite{Grant:2024ivt,Witzany:2024ttz,Mathews:2025} in the form
\begin{align}
    \mathcal{F}^K &= - \dv{K}{t} \nonumber\\
    &= \sum_{lmnk} \qty( k  - \omega_{mnk} \pdv{J_\theta}{E} - m \pdv{J_\theta}{L_z} ) \qty( \pdv{J_\theta}{K} )^{-1} Z_{lmnk} \,,
\end{align}
where $J_\theta$ is the polar action. This action and its derivatives were derived in \cite{Witzany:2024ttz}, where they are linearized in the fixed constants gauge as
\begin{align}
    J_\theta &= J_\theta^{0,\boldsymbol{C}} + s_\parallel J_\theta^{1,\boldsymbol{C}} \,, \\
    \pdv{J_\theta}{C_i} &= \pdv{J_\theta^{0,\boldsymbol{C}}}{C_i} + s_\parallel \pdv{J_\theta^{1,\boldsymbol{C}}}{C_i} \,.
\end{align}

To linearize the fluxes in the fixed $\bTC$ gauge as
\begin{equation}
    \mathcal{F}^{C} = \qty(\mathcal{F}^{C})^{0,\bTC}(\bTC) + s_\parallel \qty(\mathcal{F}^{C})^{1,\bTC}(\bTC) \,,
\end{equation}
the derivatives of the action must be transformed to this gauge. 

Let us define the auxiliary functions
\begin{subequations}\label{eq:Xi_z}
\begin{align}
    \Xi_z &\equiv \qty( 2 \pdv{J_\theta}{K} )^{-1} \,, \\
    \Xi_{tz} &\equiv \pdv{J_\theta}{E} \qty( 2 \pdv{J_\theta}{K} )^{-1} \,, \\
    \Xi_{\phi z} &\equiv - \pdv{J_\theta}{L_z} \qty( 2 \pdv{J_\theta}{K} )^{-1} \,. 
\end{align}
\end{subequations}
The evolution of the Carter constant is then written as
\begin{align}
    \mathcal{F}^K &= \sum_{lmnk} 2 \qty( k \Xi_z - \omega_{mnk} \Xi_{tz} + m \Xi_{\phi z} ) Z_{lmnk} \,.
\end{align}
The geodesic parts of the quantities \eqref{eq:Xi_z} are simply $\Upsilon_z^0$, $\Upsilon_{tz}^0$, and $\Upsilon_{\phi z}^0$, respectively. By using Eq.~\eqref{eq:transformation_Ctilde}, the linear parts in the fixed $\bTC$ gauge can be expressed as
\begin{subequations}\label{eq:Xi_corrections}
\begin{align}
    \Xi_z^{1,\bTC} &= \frac{\Tilde{E} \Upsilon_z^{0}}{2 \sqrt{{\Tilde{K}}}} \,, \\
    \Xi_{t z}^{1,\bTC} &= \frac{\Tilde{E} \Upsilon_{tz}^{0} - 3 \Upsilon_{\tau z}^{0} - a^2}{2 \sqrt{{\Tilde{K}}}} \,, 
\end{align}
\begin{align}
    \Xi_{\phi z}^{1,\bTC} &= \frac{\Tilde{E} \Upsilon_{\phi z}^{0} - a}{2 \sqrt{{\Tilde{K}}}} \,.
\end{align}
\end{subequations}
While the relation for $\Xi_z^{1,\bTC}$ can be checked analytically, we checked the relations for $\Xi_{tz}^{1,\bTC}$ and $\Xi_{\phi z}^{1,\bTC}$ numerically across the parameter space due to their complexity. By using these relations, the flux of the Carter constant can be expressed as
\begin{widetext}
\begin{equation}
    \mathcal{F}^K = \sum_{lmnk} \qty( \qty( 2 + s_\parallel \frac{\Tilde{E}}{\sqrt{\Tilde{K}}} )\qty( k \Upsilon_z - \omega_{mnk} \Upsilon_{tz} + m \Upsilon_{\phi z} ) - \frac{a s_\parallel}{\sqrt{\Tilde{K}}} \qty( m - a\omega_{mnk} ) ) Z_{lmnk} \,,
\end{equation}
where $\Upsilon_z$ is the Mino-time polar frequency and
\begin{align}
    \Upsilon_{tz}  &= \Upsilon_{tz}^0 - s_\parallel \frac{3 \Upsilon_{\tau z}^0}{2\sqrt{\Tilde{K}}} \, , \\
    \Upsilon_{\phi z} &= \Upsilon_{\phi z}^0 \,.
\end{align}

The energy and angular momentum fluxes in Eqs.~\eqref{eq:fluxes_En_Lz} can be easily linearized. By using the relations \eqref{eq:Xi_corrections}, the linear parts of the fluxes in the fixed $\bTC$ gauge can be written as
\begin{subequations}\label{eq:fluxes_constants_linear}
\begin{align}
    \mathcal{F}^{E,1} &= \sum_{lmnk} \qty( \omega_{mnk}^{1} Z_{lmnk}^0 + \omega_{mnk}^0 Z_{lmnk}^1 ) \,, \\
    \mathcal{F}^{L_z,1} &= \sum_{lmnk} m Z_{lmnk}^1 \,, \\
    \mathcal{F}^{K,1} &= \sum_{lmnk} \Bigg( \frac{\Tilde{E} \qty( k \Upsilon_z^0  - \omega_{mnk}^0 \Upsilon_{tz}^0 + m \Upsilon_{\phi z}^0 ) - a ( m - a \omega_{mnk}^0 ) - 3 \omega_{mnk}^0 \frac{ \Upsilon_ {\tau r}^0 \Upsilon_{t z}^0 - \Upsilon_{\tau z}^0 \Upsilon_{t r}^0 }{\Upsilon_t^0} }{\sqrt{\Tilde{K}}} Z_{lmnk}^0 \nonumber\\
    &\phantom{=} + 2 \qty( k \Upsilon_z^0  - \omega_{mnk}^0 \Upsilon_{tz}^0 + m \Upsilon_{\phi z}^0 ) Z_{lmnk}^1 \Bigg) \,,
\end{align}
\end{subequations}
where the linear part of $Z_{lmnk}$ is 
\begin{align}
    Z_{lmnk}^{1} &= \frac{ 2 \Re{C^{+,1}_{lmnk} \overline{C^{+,0}_{lmnk}}} + 2 \alpha_{lmnk}^0 \Re{C^{-,1}_{lmnk} \overline{C^{+,0}_{lmnk}}} + \omega_{mnk}^1 \partial_\omega \alpha_{lmnk}^{0} \abs{C^{-,0}_{lmnk}}^2}{4\pi \qty(\omega_{mnk}^0)^3} - \frac{3 \omega_{mnk}^{1}}{\omega_{mnk}^0} Z^0_{lmnk} \,
\end{align}
\end{widetext}
where
\begin{equation}
    \omega_{mnk}^1 = \omega_{mnk}^0 \frac{3 \Upsilon_\tau^0}{2 \Upsilon_t^0 \sqrt{\Tilde{K}}} \,.
\end{equation}

\begin{table*}
    \begin{tabular}{r|r|r|r|r|r|r|r|r|r}
        \thead{$n$} & \thead{$k$} & \thead{$\mathcal{F}^{E,+,1}_{lmnk}$} & \thead{$\Delta \mathcal{F}^{E,+,1}_{lmnk}$} & \thead{$\delta \mathcal{F}^{E,+,1}_{lmnk}$} & \thead{$\mathcal{F}^{E,-,1}_{lmnk}$} & \thead{$\Delta \mathcal{F}^{E,-,1}_{lmnk}$} & \thead{$\delta \mathcal{F}^{E,-,1}_{lmnk}$} \\\hline
0 & -1 & $8.32895 \times 10^{-13}$ & $2.7 \times 10^{-21}$ & $3.2 \times 10^{-9\phantom{0}}$ & $-8.74304 \times 10^{-18}$ & $3.4 \times 10^{-27}$ & $3.9 \times 10^{-10}$ \\
0 & 0 & $-2.04616 \times 10^{-11}$ & $3.4 \times 10^{-19}$ & $1.7 \times 10^{-8\phantom{0}}$ & $1.40840 \times 10^{-17}$ & $4.7 \times 10^{-26}$ & $3.3 \times 10^{-9\phantom{0}}$ \\
0 & 1 & $1.20171 \times 10^{-9\phantom{0}}$ & $7.0 \times 10^{-17}$ & $5.8 \times 10^{-8\phantom{0}}$ & $-2.12998 \times 10^{-16}$ & $1.5 \times 10^{-24}$ & $7.2 \times 10^{-9\phantom{0}}$ \\
0 & 2 & $-1.01138 \times 10^{-13}$ & $9.0 \times 10^{-20}$ & $8.9 \times 10^{-7\phantom{0}}$ & $1.28098 \times 10^{-18}$ & $4.4 \times 10^{-27}$ & $3.4 \times 10^{-9\phantom{0}}$ \\
4 & -1 & $5.09554 \times 10^{-12}$ & $6.5 \times 10^{-17}$ & $1.3 \times 10^{-5\phantom{0}}$ & $-4.19246 \times 10^{-16}$ & $2.8 \times 10^{-24}$ & $6.7 \times 10^{-9\phantom{0}}$ \\
4 & 0 & $-4.49048 \times 10^{-10}$ & $2.8 \times 10^{-16}$ & $6.1 \times 10^{-7\phantom{0}}$ & $2.41712 \times 10^{-15}$ & $5.0 \times 10^{-24}$ & $2.1 \times 10^{-9\phantom{0}}$ \\
4 & 1 & $9.07090 \times 10^{-9\phantom{0}}$ & $1.5 \times 10^{-14}$ & $1.7 \times 10^{-6\phantom{0}}$ & $-3.34835 \times 10^{-15}$ & $2.9 \times 10^{-23}$ & $8.5 \times 10^{-9\phantom{0}}$ \\
4 & 2 & $-1.29001 \times 10^{-12}$ & $7.9 \times 10^{-19}$ & $6.1 \times 10^{-7\phantom{0}}$ & $5.32186 \times 10^{-18}$ & $2.5 \times 10^{-25}$ & $4.6 \times 10^{-8\phantom{0}}$ \\
8 & -1 & $2.06516 \times 10^{-11}$ & $2.6 \times 10^{-16}$ & $1.3 \times 10^{-5\phantom{0}}$ & $-7.31491 \times 10^{-15}$ & $1.0 \times 10^{-23}$ & $1.4 \times 10^{-9\phantom{0}}$ \\
8 & 0 & $2.17149 \times 10^{-9\phantom{0}}$ & $1.8 \times 10^{-15}$ & $8.1 \times 10^{-7\phantom{0}}$ & $-1.82083 \times 10^{-14}$ & $8.6 \times 10^{-24}$ & $4.7 \times 10^{-10}$ \\
8 & 1 & $-2.25480 \times 10^{-8\phantom{0}}$ & $5.8 \times 10^{-14}$ & $2.6 \times 10^{-6\phantom{0}}$ & $8.45189 \times 10^{-15}$ & $7.5 \times 10^{-24}$ & $8.9 \times 10^{-10}$ \\
8 & 2 & $-1.97802 \times 10^{-11}$ & $4.6 \times 10^{-18}$ & $2.3 \times 10^{-7\phantom{0}}$ & $4.39762 \times 10^{-16}$ & $4.5 \times 10^{-24}$ & $1.0 \times 10^{-8\phantom{0}}$ \\
12 & -1 & $5.86402 \times 10^{-12}$ & $2.0 \times 10^{-16}$ & $3.4 \times 10^{-5\phantom{0}}$ & $-2.69434 \times 10^{-15}$ & $8.2 \times 10^{-25}$ & $3.0 \times 10^{-10}$ \\
12 & 0 & $2.89536 \times 10^{-9\phantom{0}}$ & $1.6 \times 10^{-15}$ & $5.4 \times 10^{-7\phantom{0}}$ & $-2.49660 \times 10^{-14}$ & $4.6 \times 10^{-23}$ & $1.9 \times 10^{-9\phantom{0}}$ \\
12 & 1 & $8.98065 \times 10^{-8\phantom{0}}$ & $1.1 \times 10^{-13}$ & $1.2 \times 10^{-6\phantom{0}}$ & $-8.19945 \times 10^{-14}$ & $1.1 \times 10^{-22}$ & $1.4 \times 10^{-9\phantom{0}}$ \\
12 & 2 & $4.34277 \times 10^{-11}$ & $6.9 \times 10^{-17}$ & $1.6 \times 10^{-6\phantom{0}}$ & $-9.65563 \times 10^{-16}$ & $1.2 \times 10^{-23}$ & $1.3 \times 10^{-8\phantom{0}}$ \\
16 & -1 & $6.62368 \times 10^{-13}$ & $7.4 \times 10^{-18}$ & $1.1 \times 10^{-5\phantom{0}}$ & $-2.89210 \times 10^{-16}$ & $2.5 \times 10^{-24}$ & $8.5 \times 10^{-9\phantom{0}}$ \\
16 & 0 & $6.10089 \times 10^{-10}$ & $6.6 \times 10^{-15}$ & $1.1 \times 10^{-5\phantom{0}}$ & $-6.26873 \times 10^{-15}$ & $9.4 \times 10^{-23}$ & $1.5 \times 10^{-8\phantom{0}}$ \\
16 & 1 & $3.52859 \times 10^{-8\phantom{0}}$ & $3.5 \times 10^{-14}$ & $1.0 \times 10^{-6\phantom{0}}$ & $-4.33807 \times 10^{-14}$ & $2.0 \times 10^{-23}$ & $4.7 \times 10^{-10}$ \\
16 & 2 & $2.07097 \times 10^{-10}$ & $1.7 \times 10^{-15}$ & $8.0 \times 10^{-6\phantom{0}}$ & $-2.38801 \times 10^{-15}$ & $1.6 \times 10^{-23}$ & $6.9 \times 10^{-9\phantom{0}}$ \\
20 & -1 & $3.98604 \times 10^{-14}$ & $1.4 \times 10^{-17}$ & $3.6 \times 10^{-4\phantom{0}}$ & $-1.50083 \times 10^{-17}$ & $1.4 \times 10^{-24}$ & $9.2 \times 10^{-8\phantom{0}}$ \\
20 & 0 & $5.33100 \times 10^{-11}$ & $1.7 \times 10^{-15}$ & $3.2 \times 10^{-5\phantom{0}}$ & $-6.49980 \times 10^{-16}$ & $1.7 \times 10^{-23}$ & $2.7 \times 10^{-8\phantom{0}}$ \\
20 & 1 & $4.51477 \times 10^{-9\phantom{0}}$ & $1.1 \times 10^{-14}$ & $2.5 \times 10^{-6\phantom{0}}$ & $-7.12076 \times 10^{-15}$ & $8.5 \times 10^{-22}$ & $1.2 \times 10^{-7\phantom{0}}$ \\
20 & 2 & $9.60635 \times 10^{-11}$ & $8.7 \times 10^{-16}$ & $9.0 \times 10^{-6\phantom{0}}$ & $-8.33516 \times 10^{-16}$ & $1.3 \times 10^{-24}$ & $1.5 \times 10^{-9\phantom{0}}$
    \end{tabular}
    \caption{Linear parts of the energy flux to infinity (third column) and to the horizon (sixth column) in the fixed $(E, L_z, K - a s_\parallel \sgn(L_z - a E))$ gauge for $a=0.95M$, $p=10$, $e=0.5$, $x=0.5$, $l=5$, $m=4$, and various $n$ and $k$ (first and second column, respectively). Fourth and seventh (fifth and eighth) columns show absolute (relative) differences between the values obtained using \texttt{KerrSpinningFluxes} and \texttt{Spinning-Body-Hamilton-Jacobi} \cite{Piovano:2024}.}
    \label{tab:rel_errors}
\end{table*}

The new implementation of Teukolsky amplitudes and fluxes is validated by calculating the relative and absolute differences of the linear parts of energy fluxes $\mathcal{F}^{E,1,\bC}$ for certain mode numbers calculated using our code and the code \texttt{Spinning-Body-Hamilton-Jacobi} by \citet{Piovano:2024}. The validation was done as follows. First we calculated the fluxes using \texttt{KerrSpinningFluxes} for several values of spin with fixed constants $(E, L_z, K - 2 a s_\parallel \sgn(L_z - a E))$. Then we calculated the linear-in-spin parts using fourth-order finite difference formula with step $\Delta s_\parallel/M = 10^{-3}$. The truncation error should therefore be $\order{10^{-12}}$. Then, the linear parts in the same gauge were computed directly using \texttt{Spinning-Body-Hamilton-Jacobi}. In both cases the input precision for the trajectories was set to 45 digits of precision and we used 16 Fourier coefficients for the trajectory corrections in \texttt{Spinning-Body-Hamilton-Jacobi}. From the results presented in Table \ref{tab:rel_errors} we can see that the relative error is at most $3.6 \times 10^{-4}$ for the infinity flux with $n=20, k=-1$. However, this is a subdominant mode and the relative error for this mode decreases to $5.9\times 10^{-6}$ when we increase the initial precision of \texttt{Spinning-Body-Hamilton-Jacobi} to 60 digits.

\begin{figure}
    \centering
    \includegraphics[width=\linewidth]{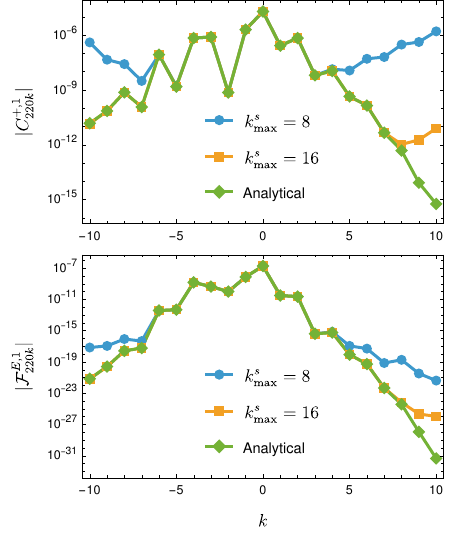}
    \caption{Linear parts of the infinity amplitudes (top) and total energy fluxes (bottom) calculated using the analytical trajectory and the trajectory calculated using the Drummond and Hughes method with different numbers of Fourier modes $k^s_{\text{max}}$. $a=0.9M$, $p=12$, $e=0.2$, $x=0.5$, $l=2$, $m=2$, $n=0$.}
    \label{fig:Clmnk_k}
\end{figure}

\begin{figure}
    \centering
    \includegraphics[width=\linewidth]{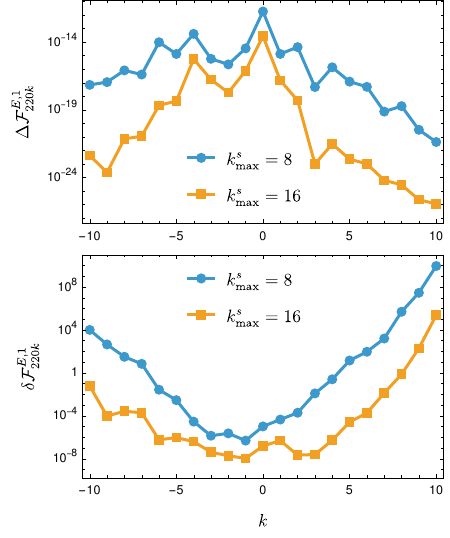}
    \caption{Absolute (top) and relative (bottom) difference of linear parts of energy fluxes. $a=0.9M$, $p=12$, $e=0.2$, $x=0.5$, $l=2$, $m=2$, $n=0$.}
    \label{fig:DeltaFlmnk}
\end{figure}

To demonstrate the advantages of the new analytical calculations, we plot the linear parts of the amplitudes and energy fluxes calculated using the analytic trajectory and the trajectory computed in \cite{Drummond:2022efc} against the mode number $k$ in Fig.~\ref{fig:Clmnk_k}. Figure~\ref{fig:DeltaFlmnk} illustrates the absolute and relative differences in the linear parts of the energy flux in fixed $p,e,x$ gauge calculated with the analytical trajectory compared to calculation using the Drummond and Hughes trajectory. We can see that, with the analytical trajectory, the modes of the flux converge even for higher $k$ modes, while for the trajectory by Drummond and Hughes, the accuracy is limited by the number of Fourier modes included in the trajectory $k^s_\text{max}$ and deteriorates quickly with increasing $\abs{k}$. We set the initial precision for the Drummond and Hughes trajectory to 64 digits of precision while the analytical trajectory was calculated using 32 digits of precision. To extract the linear-in-spin parts we used fourth order finite-difference formulas with step size $\Delta s_\parallel/M = 10^{-2}$. 

Additionally, our calculation, which utilizes a separable analytical solution, is faster than the calculation from \cite{Drummond:2022efc,Piovano:2024} which uses two-parametric Fourier series and requires choosing a balance between speed and accuracy \cite{Drummond:2026}. For illustration, the $l=5$, $m=4$, $n=12$, $k=1$ mode from Table \ref{tab:rel_errors} needs precalculating the trajectory and its linear part which takes approximately $0.05$ seconds ($60$ seconds) in \texttt{KerrSpinningFluxes} (in \texttt{Spinning-Body-Hamilton-Jacobi}). During the integration of the Teukolsky source term, the trajectory and its correction must be evaluated on a grid in $q_r$ and $q_z$ (here with 656 and 48 steps, respectively). This takes approximately 1 second (46 seconds) in \texttt{KerrSpinningFluxes} (in \texttt{Spinning-Body-Hamilton-Jacobi}).

\section{Flux-driven inspirals}\label{sec:inspirals}

The fluxes $\mathcal{F}^E, \mathcal{F}^{L_z}, \mathcal{F}^K$ derived in the previous section can now be utilized in Eqs.~\eqref{eq:multiscale_evolution} to calculate the inspirals from the flux-balance laws
\begin{align}
    \dv{E}{t} &= F^E = -\mathcal{F}^E \,, \\
    \dv{L_z}{t} &= F^{L_z} = -\mathcal{F}^{L_z} \,, \\
    \dv{K}{t} &= F^K = -\mathcal{F}^K \,.
\end{align}
Generally, the evolution of any set of parameters $\pi_i$ is determined by the rates of change of the constants and the Jacobian matrix as
\begin{equation}
    \dv{\pi_i}{t} = F^{\pi}_i = \pdv{\pi_i}{C_j} F^{C}_j \,.
\end{equation}
After the expansion in the secondary spin, this equation reads
\begin{multline}
    \dv{\pi_i}{t} = \qty( \pdv{\pi_i^0}{C_j} + s_\parallel \pdv{\pi^{1,\bC}}{C_j} ) (F^{C,0}_j + s_\parallel F^{C,1,\boldsymbol{\pi}}_j) \\
    = \pdv{\pi_i^0}{C_j} F^{C,0}_j + s_\parallel \qty( \pdv{\pi_i^0}{C_j} F^{C,1,\boldsymbol{\pi}}_j + \pdv{\pi^{1,\bC}}{C_j} F^{C,0}_j ) \,.
\end{multline}
If we consider $\boldsymbol{\pi} = \bTC$, the evolution of the shifted constants reads
\begin{multline}\label{eq:dCTildedt}
    \dv{\Tilde{C}_i}{t} = F^{C,0}_i + s_\parallel \qty( F^{C,1,\bTC}_i + \pdv{\Tilde{C}^{1,\bC}_i}{C_j} F^{C,0}_j ) \,.
\end{multline}
This form is advantageous because the derivatives $\pdv*{\Tilde{C}^{1,\bC}_i}{C_j} = - \pdv*{C^{1,\bTC}_i}{C_j}$ can be easily inferred from Eqs.~\eqref{eq:constants_corrections}.

In practice, it is convenient to evolve the virtual-worldline parameters $(\Tilde{p},\Tilde{e},\Tilde{x})$ which are geodesically related to the shifted constants. Their evolution then reads
\begin{subequations}\label{eq:evolution_pTilde}
\begin{align}
    \dv{\Tilde{p}}{t} &= \pdv{p^0}{C_j} F^{C,0}_j + s_\parallel \pdv{p^0}{C_j} \qty(F^{C,1,\boldsymbol{\pi}}_j + \pdv{\Tilde{C}^{1,\bC}_j}{C_k} F^{C,0}_k) \,, \\
    \dv{\Tilde{e}}{t} &= \pdv{e^0}{C_j} F^{C,0}_j + s_\parallel \pdv{e^0}{C_j} \qty(F^{C,1,\boldsymbol{\pi}}_j + \pdv{\Tilde{C}^{1,\bC}_j}{C_k} F^{C,0}_k) \,, \\
    \dv{\Tilde{x}}{t} &= \pdv{x^0}{C_j} F^{C,0}_j + s_\parallel \pdv{x^0}{C_j} \qty(F^{C,1,\boldsymbol{\pi}}_j + \pdv{\Tilde{C}^{1,\bC}_j}{C_k} F^{C,0}_k) \,.
\end{align}
\end{subequations}
In these expressions, we can clearly see the linear-in-spin corrections to the forcing functions introduced in Eqs.~\eqref{eq:multiscale_evolution} and \eqref{eq:1PA_functions}. The set of evolution equations for the inspiral also includes the evolution of the phases 
\begin{equation}\label{eq:evolution_Phi}
    \dv{\Phi_j}{t} = \Omega_{j}^0 + s_\parallel \Omega_j^{1,\bTC} = \Omega_j^0 \qty( 1 + s_\parallel \frac{3 \Upsilon_\tau^0}{2 \Upsilon_t^0 \sqrt{\Tilde{K}}} ) \,,
\end{equation}
where we utilized the linear parts from Eq.~\eqref{eq:frequencies_corrections}.

\subsection{Consistency check for equatorial orbits}

Due to the properties of the geodesic Carter constant flux, the rate of change of the inclination parameter $x$ vanishes for equatorial orbits, i.e., equatorial orbits remain equatorial \cite{Sago:2006}. The same behavior can be proven for the linear-in-spin part of the flux as follows. First, we consider the flux of the virtual Carter constant
\begin{equation}
    \Tilde{Q} = \Tilde{K} - (\Tilde{L}_z - a \Tilde{E})^2 = Q + s_\parallel \frac{2 a (L_z - a E) - Q E}{\sqrt{Q + (L_z-aE)^2}} \,.
\end{equation}
This constant approaches zero for equatorial orbits, as can be observed from the fact that
\begin{equation}
    Q \rightarrow -2 a s_\parallel \sgn(L_z-a E)
\end{equation}
in the equatorial limit \cite{Tanaka:1996ht,Witzany:2019}. 

The rate of change of $\Tilde{Q}$ can be derived from Eqs.~\eqref{eq:fluxes_constants_linear} and \eqref{eq:dCTildedt} in the form
\begin{multline}\label{eq:dQTdt}
    \dv{\Tilde{Q}}{t} = - \sum_{lmnk} 2 \qty(k \Upsilon_z + m \mathcal{C}_m - \omega_{mnk} \mathcal{C}_\omega ) \\ \times \qty( 1 - s_\parallel \frac{\qty(\Tilde{L}_z - a \Tilde{E}) \qty( 2 a + \Tilde{E} \qty(\Tilde{L}_z-a\Tilde{E}) )}{2 \Tilde{K}^{3/2}} ) Z_{lmnk} \,,
\end{multline}
where
\begin{align}
    \mathcal{C}_m &= \Tilde{L}_z \expval{\cot^2\Tilde{\theta}} + s_\parallel \Tilde{Q} \frac{\Tilde{E} (\Tilde{L}_z-a \Tilde{E}) + 2 a}{2 \Tilde{K}^{3/2}} \,, \\
    \mathcal{C}_\omega &= a^2 \expval{\cos^2\Tilde{\theta}} \qty( \Tilde{E} - s_\parallel \frac{3}{2 \sqrt{\Tilde{K}}} ) \nonumber\\ &\phantom{=} + s_\parallel \Tilde{Q} \frac{\Tilde{Q} + \Tilde{L}_z^2 - a \Tilde{E} \Tilde{L}_z + 2 a^2}{2 \Tilde{K}^{3/2}} \,.
\end{align}    
Eq.~\eqref{eq:dQTdt} reduces to Eq.~(3.26) in \cite{Sago:2006} and Eq.~(3.28) in \cite{Hughes:2021} for $s_\parallel = 0$ and vanishes in the equatorial limit $\Tilde{Q} = 0$, since the averages of $\cos^2\Tilde{\theta}$ and $\cot^2\Tilde{\theta}$ and amplitudes with $k\neq 0$ also vanish.

\subsection{Nearly spherical inspirals}

In this section we present the calculation of nearly spherical inspirals using the fixed $\bTC$ gauge. Here, we utilize the fluxes generated in \cite{Skoupy:2025b}, where the geodesic part of the fluxes was calculated alongside the linear part in the fixed turning points gauge, along with the derivatives of the geodesic fluxes with respect to $p$ and $x$. This allows us to use Eq.~\eqref{eq:transformation_ptilde} to find the linear parts of the energy and angular momentum flux in the virtual-worldline gauge $\mathcal{F}^{1,\Tilde{C}}$. The inspirals can then be computed from the evolution equations for $\Tilde{p}$, $\Tilde{x}$, $\Phi_z$, and $\Phi_\phi$ \eqref{eq:evolution_pTilde} and \eqref{eq:evolution_Phi}. Additionally, we linearize the evolved quantities as
\begin{align}
    \Tilde{p}(t) &= \Tilde{p}^0(t) + s_\parallel \Tilde{p}^1(t) \,, \\
    \Tilde{x}(t) &= \Tilde{x}^0(t) + s_\parallel \Tilde{x}^1(t) \,, \\
    \Phi_z(t) &= \Phi_z^0(t) + s_\parallel \Phi_z^1(t) \,, \\
    \Phi_\phi(t) &= \Phi_\phi^0(t) + s_\parallel \Phi_\phi^1(t) 
\end{align}
to align with the results of \cite{Skoupy:2025b}.

\begin{figure}
    \centering
    \includegraphics[width=\linewidth]{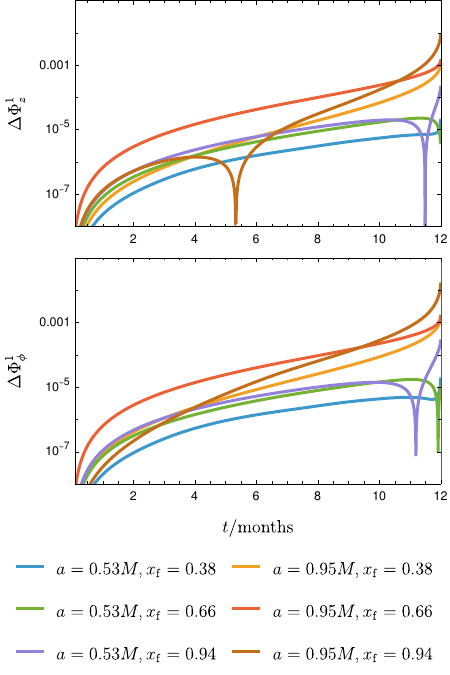}
    \caption{Phase differences between inspirals calculated in the fixed $\bTC$ gauge and fixed turning points gauge from \cite{Skoupy:2025b}.}
    \label{fig:Delta_Omega1}
\end{figure}

By using the same numerical methods described in Sec.~V in \cite{Skoupy:2025b}, we calculated the evolution of the adiabatic orbital parameters and phases $\Tilde{p}^0$, $\Tilde{x}^0$, $\Phi_{z,\phi}^0$, as well as their linear corrections $\Tilde{p}^1$, $\Tilde{x}^1$, $\Phi_{z,\phi}^1$. While the evolution of $\Tilde{p}^1$ and $\Tilde{x}^1$ differs from the results of \cite{Skoupy:2025b} due to different spin gauges, the evolution of the gauge invariant phase shifts $\Phi_{z,\phi}^1$ agrees within numerical error\footnote{This numerical error is mostly caused by the interpolation of the forcing functions. This was found by changing the grid resolution and verifying that the differences converge to zero with increasing number of grid points.}. Note that only the linear-in-spin parts should agree up to numerical error; the total nonlinearized phases and waveforms would differ at subleading orders. We plot the absolute difference between the phase shifts in Fig.~\ref{fig:Delta_Omega1}. We can see that the error peaks before the end of the inspiral at around $10^{-2}$ rad.

\section{Conclusions and summary} \label{sec:Concl}

In this paper, we utilized the analytical solution for the motion of spinning particles in Kerr spacetime \cite{Skoupy:2025} for the calculation of extreme-mass-ratio inspirals. First, we developed a new spin gauge based on fixing the shifted constants $\bTC$ that parametrize the analytical solution. We derived the transformation relations to other gauges and implemented them in a public \textit{Wolfram Mathematica} code \cite{KerrSpinningFluxes}. 

Then we focused on the calculation of GW fluxes using the Teukolsky equation with the spinning particle source term calculated from the analytical trajectory. Thanks to the properties of the analytical solution, which exploits the hidden symmetry, we were able to significantly simplify the calculation of Teukolsky amplitudes, making it comparably algebraically complex to the nonspinning case. Furthermore, because the source term is constructed from a separable analytical solution, its evaluation is significantly faster than the evaluation of the trajectory from \cite{Drummond:2022efc,Piovano:2024}. Thus, our method will be optimal for extreme orbits with high eccentricity and inclination.

Thanks to the recently found flux-balance laws \cite{Grant:2024ivt} and the analytical solution for the actions of spinning particle trajectories \cite{Witzany:2024ttz}, we found an explicit form of the Carter constant flux in terms of Teukolsky amplitudes and known geodesic functions. This enables practical calculations of generic eccentric and off-equatorial inspirals of spinning particles into a Kerr black hole.

Then we derived the linear-in-spin parts of the forcing functions for orbital parameters related to the virtual worldline, which can be used, for example, in waveform generation frameworks such as \texttt{FEW} \cite{Chapman-Bird:2025}. We checked the consistency of the evolution equations by performing the equatorial limit and verifying that equatorial orbits remain equatorial, even with a nonzero secondary spin. Finally, we validated the evolution framework by calculating nearly spherical inspirals and verifying the gauge invariance of the phase shifts.

The natural next step involves calculating the fluxes and their linear part on an interpolation grid in the $a$, $p$, $e$, and $x$ parameter space, as well as computing the inspirals and waveforms. This could be efficiently achieved by using packages like \texttt{FastEMRIWaveforms}. We leave this endeavor for future work.

Although the orthogonal component of spin is not observable in EMRIs, it could be relevant in intermediate-mass-ratio inspirals due to its presence in the amplitudes \cite{Mathews:2025b}. The analytical solution to the trajectory could be used to calculate the asymptotic amplitudes related to the secondary spin precession, which can be utilized in hybrid models that combine both post-Newtonian and self-force results \cite{Honet:2025}. Another possibility is to use the simplified expressions for the fluxes to calculate the analytical post-Newtonian expansion of the fluxes, as it could provide a sufficiently accurate approximation in some parts of the parameter space \cite{Burke:2023lno,Skoupy:2024}.


\begin{acknowledgments}

This work was supported by the Czech Science Foundation grant 26-23696S. The author would like to thank Josh Mathews for reading an earlier version of this work and giving useful comments. This work makes use of the \textit{Black Hole Perturbation Toolkit} \cite{BHPToolkit}, specifically the \texttt{KerrGeodesics} \cite{KerrGeodesicsZenodo}, \texttt{Teukolsky} \cite{TeukolskyZenodo}, and \texttt{GeneralRelativityTensors} packages.

\end{acknowledgments}

\section*{Data availability}

The data that support the findings of this article are openly available \cite{KerrSpinningFluxes}.

\appendix

\section{Jacobians}\label{app:jacobians}

In this appendix, we describe the construction of Jacobians between various orbital parameters, namely the parameters $(p,e,x)$, the constants of motion $(E,L_z,K)$, the orbital frequencies $(\Omega_r,\Omega_z,\Omega_\phi)$, and the actions $(J_r,J_\theta,J_\phi)$. Here, we consider only geodesic quantities.

\subsection{Turning points and constants}

To find the derivatives of the turning points $x_{\text{t}i} \equiv (r_1, r_2, z_1)$ with respect to the constants of motion $(E,L_z,K)$, we can use the conditions for the turning points \eqref{eq:conditions_turning_points}. By taking the derivatives of these equations with respect to the constants, we can easily express the desired derivatives of the turning points as
\begin{align}
    \pdv{r_{1,2}}{C_j} &= - \eval{\pdv{R}{C_j} \qty( \pdv{R}{r} )^{-1} }_{r = r_{1,2}} \,,  \\
    \pdv{z_1}{C_j} &= - \eval{\pdv{Z}{C_j} \qty( \pdv{Z}{z} )^{-1} }_{z = z_1} \,. 
\end{align}    
The Jacobian between the orbital parameters $p_i = (p,e,x)$ and the constants can then be expressed as
\begin{equation}
    \pdv{p_i}{C_k} = \pdv{p_i}{x_{\text{t}j}} \pdv{x_{\text{t}j}}{C_k} \,,
\end{equation}
where the derivatives $\pdv{p_i}{x_{\text{t}j}}$ in terms of $p$, $e$, and $x$ are 
\begin{subequations}\label{eq:Jacobian_pex_tp}
\begin{align}
    \pdv{p}{r_1} &= (1-e)^2/2 \,, & \pdv{p}{r_2} &= (1+e)^2/2 \,, \\
    \pdv{e}{r_1} &= \frac{(1-e^2)(1-e)}{2p} \,, & \pdv{e}{r_2} &= -\frac{(1-e^2)(1+e)}{2p} \,, \\
    \pdv{x}{z_1} &= - \frac{\sqrt{1-x^2}}{x} \,. &
\end{align}    
\end{subequations}
These expressions have already been successfully used for the adiabatic inspirals \cite{Hughes:2021}.

\subsection{Actions and constants}

Derivatives of the actions $J_r$ and $J_\theta$ are used in the formulas for orbital frequencies and are thus already known \cite{Schmidt:2002,Fujita:2009bp,vandeMeent:2020}. They can be expressed in terms of the geodesic Mino-time frequencies \eqref{eq:Mino_frequencies_geo} as
\begin{align}
    \pdv{J_r}{K} &= - \frac{1}{2 \Upsilon_r} \,,                    & \pdv{J_\theta}{K} &= \frac{1}{2 \Upsilon_z} \,, & \pdv{J_\phi}{K} &= 0 \,, \\
    \pdv{J_r}{E} &= \frac{\Upsilon_{tr}}{\Upsilon_r} \,,            & \pdv{J_\theta}{E} &= \frac{\Upsilon_{tz}}{\Upsilon_z} \,, & \pdv{J_\phi}{E} &= 0 \,, \\
    \pdv{J_r}{L_z} &= - \frac{\Upsilon_{\phi r}}{\Upsilon_r} \,,    & \pdv{J_\theta}{L_z} &= - \frac{\Upsilon_{\phi z}}{\Upsilon_z} \,, & \pdv{J_\phi}{L_z} &= 1 \,.
\end{align}

\subsection{Frequencies and constants}

The orbital frequencies, expressed using the derivatives of the actions, read \cite{Schmidt:2002,Witzany:2024ttz}
\begin{align}
    \Omega_r &= \frac{-\pdv{J_z}{K}}{ \pdv{J_z}{E} \pdv{J_r}{K} - \pdv{J_r}{E} \pdv{J_z}{K} } \,, \\
    \Omega_z &= \frac{\pdv{J_r}{K}}{ \pdv{J_z}{E} \pdv{J_r}{K} - \pdv{J_r}{E} \pdv{J_z}{K} } \,, \\
    \Omega_\phi &= \frac{ \pdv{J_r}{L_z} \pdv{J_z}{K} - \pdv{J_z}{L_z} \pdv{J_r}{K} }{ \pdv{J_z}{E} \pdv{J_r}{K} - \pdv{J_r}{E} \pdv{J_z}{K} } \,.
\end{align}
We can simply take the derivatives of these expressions with respect to $\bC$ to find $\pdv*{\bOm}{\bC}$. For these derivatives, the second derivatives of the actions are needed.

In \cite{Witzany:2024ttz}, the actions and their first derivatives were calculated. When the defining integrals are understood as integrals over a loop in phase space, i.e.,
\begin{align}
    J_r^0 = \frac{1}{2\pi} \oint \frac{\sqrt{R(r)}}{\Delta} \dd r \,,  \\
    J_z^0 = \frac{1}{2\pi} \oint \frac{\sqrt{Z(z)}}{\Delta} \dd z \,,
\end{align}
their derivatives can be calculated as
\begin{align}
    \pdv[2]{J_r}{C_i}{C_j} &= \frac{1}{2\pi} \oint \pdv[2]{}{C_i}{C_j} \frac{\sqrt{R(r)}}{\Delta} \dd r \,, \\
    \pdv[2]{J_\theta}{C_i}{C_j} &= \frac{1}{2\pi} \oint \pdv[2]{}{C_i}{C_j} \frac{\sqrt{Z(z)}}{\Delta} \dd z \,.
\end{align}
Because these integrals include terms such as $\int_{r_2}^{r_1}((r_1-r))(r-r_2))^{-3/2} \dd r$, we must employ Hadamard regularization. Then, the derivatives of the radial action can be written as
\begin{widetext}
\begin{align}
    \pdv[2]{J_r}{K} &= -\frac{1}{4\pi} {\rm Pf} \int_{r_2}^{r_1} \frac{\Delta}{(R(r))^{3/2}} \dd r \,, &
    \pdv[2]{J_r}{K}{E} &= \frac{1}{2\pi} {\rm Pf} \int_{r_2}^{r_1} \frac{(r^2+a^2)P_r}{(R(r))^{3/2}} \dd r \,, \\
    \pdv[2]{J_r}{K}{L_z} &= -\frac{a}{2\pi} {\rm Pf} \int_{r_2}^{r_1} \frac{P_r}{(R(r))^{3/2}} \dd r \,, &
    \pdv[2]{J_r}{E} &= -\frac{1}{\pi} {\rm Pf} \int_{r_2}^{r_1} \frac{(r^2+a^2)^2 (K + r^2)}{(R(r))^{3/2}} \dd r \,, \\
    \pdv[2]{J_r}{E}{L_z} &= \frac{a}{\pi} {\rm Pf} \int_{r_2}^{r_1} \frac{(r^2+a^2)(K + r^2)}{(R(r))^{3/2}} \dd r \,, &
    \pdv[2]{J_r}{L_z} &= -\frac{a^2}{4\pi} {\rm Pf} \int_{r_2}^{r_1} \frac{K + r^2}{(R(r))^{3/2}} \dd r \,,
\end{align}
and similar for the polar action derivatives. After applying the regularization techniques from \cite{Witzany:2024ttz}, the radial action derivatives are
\begin{align}
    \pdv[2]{J_r}{K} &= \frac{\mathcal{C}}{4 \pi (1-E^2)} \sum_{i=1}^4 \Delta(r_i) \frac{I^r_i}{\Pi_i} \,, \\
    \pdv[2]{J_r}{K}{E} &= -\frac{\mathcal{C}}{2 \pi (1-E^2)} \qty( E \elK(k_r) + \sum_{i=1}^4 (r_i^2+a^2)((r_i^2+a^2) E - a L_z) \frac{I^r_i}{\Pi_i} ) \,, \\
    \pdv[2]{J_r}{K}{L_z} &= \frac{a \mathcal{C}}{2 \pi (1-E^2)} \sum_{i=1}^4 ((r_i^2+a^2) E - a L_z) \frac{I^r_i}{\Pi_i} \,, \\
    \pdv[2]{J_r}{E} &= \frac{\mathcal{C}}{\pi (1-E^2)} \qty( \frac{4 - (1-E^2)(E^2 K + 2 a E L_z - a^2)}{(1-E^2)^2} \elK(k_r) + \frac{2}{1-E^2} I_r + I_{r^2} + \sum_{i=1}^4 (r_i^2+a^2)^2(K + r_i) \frac{I^r_i}{\Pi_i} ) \,, \\
    \pdv[2]{J_r}{E}{L_z} &= \frac{a\mathcal{C}}{\pi (1-E^2)} \qty( - \elK(k_r) + \sum_{i=1}^4 (r_i^2+a^2)(K+r_i^2) \frac{I^r_i}{\Pi_i} ) \,, \\
    \pdv[2]{J_r}{L_z} &= \frac{a^2 \mathcal{C}}{2 \pi (1-E^2)} \sum_{i=1}^4 (K + r_i^2) \frac{I^r_i}{\Pi_i} \,, 
\end{align}
where
\begin{align}
    \mathcal{C} &= \frac{2}{\sqrt{(1-E^2)(r_1-r_3)(r_2-r_4)}} \,, \\
    k_r^2 &= \frac{(r_1-r_2)(r_3-r_4)}{(r_1-r_3)(r_2-r_4)} \,, \\
    \Pi_i &= \prod_{j=1,j\neq i}^4 (r_i - r_j) \,, \\
    I_r &= r_3 \elK(k_r) + (r_2-r_3) \elPi(h_r,k_r) \,, \\
    I_{r^2} &= \frac{1}{2} ( \qty( r_3 (r_2+r_3) - r_1 (r_3-r_2) ) \elK(k_r) + (r_1-r_3)(r_2-r_4) \elE(k_r) + (r_2-r_3)(r_1+r_2+r_3+r_4) \elPi(h_r,k_r)) \,, \\
    h_r &= \frac{r_1-r_2}{r_1-r_3} \,, 
\end{align}
\begin{align}
    I_1^r &= \frac{(r_2-r_4) \elE(k_r) - (r_1-r_4) \elK(k_r)}{(r_1-r_2)(r_1-r_4)} \,, & 
    I_2^r &= \frac{-(r_1-r_3) \elE(k_r) + (r_2-r_3) \elK(k_r)}{(r_1-r_2)(r_2-r_3)} \,,
    \\
    I_3^r &= \frac{(r_2-r_4) \elE(k_r) - (r_2-r_3) \elK(k_r)}{(r_2-r_3)(r_3-r_4)} \,, &
    I_4^r &= \frac{-(r_1-r_3) \elE(k_r) + (r_1-r_4) \elK(k_r)}{(r_3-r_4)(r_1-r_4)} \,.
\end{align} 

The polar action derivatives read
\begin{align}
    \pdv[2]{J_\theta}{K} &= - \frac{1}{2\pi a^3 z_2^3 (1-E^2)^{3/2} (z_2^2-z_1^2)} \sum_{i=1}^2 \qty(1-z_i^2) I_i^z \,, \\
    \pdv[2]{J_\theta}{K}{E} &= \frac{1}{\pi a z_2 (1-E^2)^{3/2}} \qty( E \elK(k_z) - \frac{1}{a z_2^2 (z_2^2-z_1^2)} \sum_{i=1}^2 \qty(1-z_i^2)\qty( L_z - a E(1-z^2) ) I_i^z ) \,, \\
    \pdv[2]{J_\theta}{K}{L_z} &= \frac{1}{\pi a^3 z_2^3 (1-E^2)^{3/2} (z_2^2-z_1^2)} \sum_{i=1}^2 \qty(L_z - a E (1-z_i^2)) I_i^z \,, \\
    \pdv[2]{J_\theta}{E} &= \frac{2}{\pi a z_2 (1-E^2)^{3/2}} \Bigg( -\qty( K - a^2 \qty(z_1^2 - 2 \qty(1-z_2^2) ) ) \elK(k_z) - a^2 z_2^2 \elE(k_z) \nonumber \\ &\phantom{=} - \frac{1}{z_2^2 \qty(z_2^2-z_1^2)} \sum_{i=1}^2 \qty(1-z_i^2)^2\qty( K - a^2 z_i^2 ) I_i^z \Bigg) \,, \\
    \pdv[2]{J_\theta}{E}{L_z} &= \frac{2}{\pi a z_2 (1-E^2)^{3/2}} \qty( a \elK(k_z) + \frac{1}{a z_2^2 (z_2^2-z_1^2)} \sum_{i=1}^2 \qty(1-z_i^2)\qty( K - a^2 z_i^2 ) I_i^z ) \,, \\
    \pdv[2]{J_\theta}{L_z} &= - \frac{2}{\pi a^3 z_2^3 (1-E^2)^{3/2} (z_2^2-z_1^2)} \sum_{i=1}^2 \qty(K - a z_i^2) I_i^z \,, 
\end{align}
where
\begin{align}
    k_z^2 &= \frac{z_1^2}{z_2^2} \,,  \\
    I^z_1 &= \frac{1}{k_z^2} \qty( \elK(k_z) - \frac{\elE(k_z)}{1-k_z^2} ) = \frac{\elD(k_z) - \elK(k_z)}{1-k_z^2} \,, \\
    I^z_2 &= - \frac{\elE(k_z)}{1-k_z^2} \,.
\end{align}
Here we used the elliptic integral of Legendre's type $\elD(k)$
\begin{equation}
    \elD(k) = \int_0^t \frac{t^2}{\sqrt{(1-t^2)(1-k^2 t^2)}} \dd t  = \frac{\elK(k)-\elE(k)}{k^2} = \frac{R_D(0,1-k^2,0)}{3} \,, 
\end{equation}
where $R_D(x,y,z)$ is Carlson's elliptic integral. The use of the set of elliptic integrals $\elK$, $\elD$, $\elPi$ regularizes the results for equatorial orbits.

\begin{figure}
    \centering
    \includegraphics[width=0.48\linewidth]{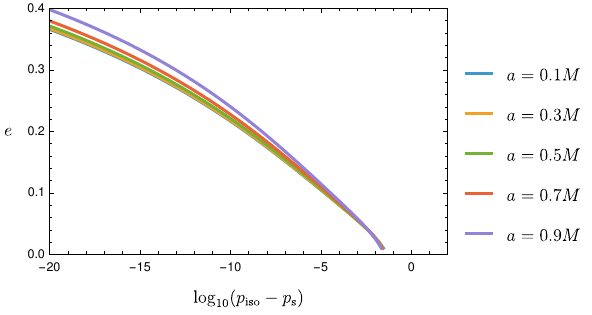} \includegraphics[width=0.48\linewidth]{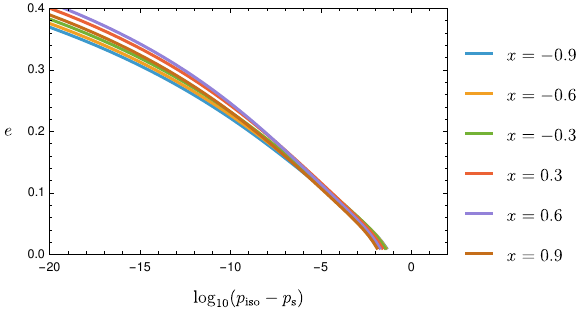} 
    \caption{Isofrequency line in the $(p,e)$ plane for different Kerr parameters $a$ and $x=0.5$ (left) and for different inclinations $x$ and $a=0.95M$ (right).}
    \label{fig:isofreq}
\end{figure}

These derivatives can be utilized for the calculation of the isofrequency line. It was found that there exist pairs of orbits with different constants of motion and orbital parameters, but with identical frequencies \cite{Warburton:2013}. These pairs are separated by a two-dimensional surface in the $(E,L_z,K)$ space, where $\det{\pdv*{\bOm}{\bC}}$ vanishes. Using our formulas, it is easy to numerically find this surface. In Fig.~\ref{fig:isofreq} we show the isofrequency line in the $(p,e)$ plane for different Kerr parameters $a$ and inclinations $x$ against the distance from the separatrix $p_{\text{s}}$. We can see that it has a maximum at $e=0$ and decreases below $10^{-20}$ at $e \approx 0.4$.

\end{widetext}

\section{Linear parts in different gauges}\label{app:linear_parts}

Here we list all relations from Eqs. \eqref{eq:gauge_transformation} and \eqref{eq:gauge_transformation2} for the linear parts of the shifted constants $\bTC = (\Tilde{E},\Tilde{L}_z,\Tilde{K})$, the constants of motion $\bC = (E,L_z,K)$, the frequencies $\bOm = (\Omega_r, \Omega_z, \Omega_\phi)$, the orbital parameters $\bp = (p,e,x)$ and the actions $\bJ = (J_r,J_\theta,J_\phi)$ from Table \ref{tab:linear_parts} in different spin gauges. All quantities are calculated from the linear parts of the constants, orbital parameters or frequencies in the fixed $\bTC$ gauge, or from the linear parts of the actions in the fixed $\bC$ gauge from \cite{Witzany:2024ttz},
\begin{align}
    \bTC^{1,\bC} &= - \bC^{1,\bTC} \,, \\
    \bTC^{1,\bp} &= - \qty( \pdv{\bp}{\bC} )^{-1} \cdot \bp^{1,\bTC} \,, \\
    \bTC^{1,\bOm} &= - \qty( \pdv{\bOm}{\bC} )^{-1} \cdot \bOm^{1,\bTC} \,, \\
    \bOm^{1,\bC} &= \bOm^{1,\bTC} - \pdv{\bOm}{\bC} \cdot \bC^{1,\bTC} \,, \\
    \bOm^{1,\bp} &= \bOm^{1,\bTC} - \pdv{\bOm}{\bC} \cdot \qty( \pdv{\bp}{\bC} )^{-1} \cdot \bp^{1,\bTC} \,, \\
    \bC^{1,\bp} &= \bC^{1,\bTC} - \qty( \pdv{\bp}{\bC} )^{-1} \cdot \bp^{1,\bTC} \,, 
\end{align}
\begin{align}
    \bC^{1,\bOm} &= \bC^{1,\bTC} - \qty( \pdv{\bOm}{\bC} )^{-1} \cdot \bOm^{1,\bTC} \,, \\
    \bp^{1,\bC} &= \bp^{1,\bTC} - \pdv{\bp}{\bC} \cdot \bC^{1,\bTC} \,,  \\
    \bp^{1,\bOm} &= \bp^{1,\bTC} - \pdv{\bp}{\bC} \cdot \qty(\pdv{\bOm}{\bC})^{-1} \cdot \bOm^{1,\bTC}\,, \\
    \bJ^{1,\bTC} &= \bJ^{1,\bC} + \pdv{\bJ}{\bC} \cdot \bC^{1,\bTC} \,, \\
    \bJ^{1,\bp} &= \bJ^{1,\bC} + \pdv{\bJ}{\bC} \cdot \qty(\bC^{1,\bTC} - \qty( \pdv{\bp}{\bC} )^{-1} \cdot \bp^{1,\bTC} ) \,, \\
    \bJ^{1,\bOm} &= \bJ^{1,\bC} + \pdv{\bJ}{\bC} \cdot \qty( \bC^{1,\bTC} - \qty( \pdv{\bOm}{\bC} )^{-1} \cdot \bOm^{1,\bTC} ) \,, \\
    \bTC^{1,\bJ} &= - \bC^{1,\bTC} - \qty(\pdv{\bJ}{\bC})^{-1} \cdot \bJ^{1,\bC} \\
    \bOm^{1,\bJ} &= \bOm^{1,\bTC} - \pdv{\bOm}{\bC} \cdot \qty( \bC^{1,\bTC} + \qty(\pdv{\bJ}{\bC})^{-1} \cdot \bJ^{1,\bC} ) \,, \\
    \bC^{1,\bJ} &= - \qty(\pdv{\bJ}{\bC})^{-1} \cdot \bJ^{1,\bC} \,, \\
    \bp^{1,\bJ} &= \bp^{1,\bTC} - \pdv{\bp}{\bC} \cdot \qty( \bC^{1,\bTC} + \qty(\pdv{\bJ}{\bC})^{-1} \cdot \bJ^{1,\bC} ) \,.
\end{align}

\begin{widetext}
\section{Teukolsky source term}\label{app:source}

In this appendix we list the expressions appearing in the Teukolsky amplitude in Eq.~\eqref{eq:Cpmlmnk_final}. In particular, the functions $F^\pm_{ab}$ and $G^\pm_a$ are
\begin{align}
    F_{nn}^\pm &= - \frac{2 \zeta^3 \bar{\zeta}}{\Delta^2} \qty( \mathcal{L}^\dag_1 \mathcal{L}^\dag_2 S - \frac{2 i a \sin\theta}{\zeta} \mathcal{L}^\dag_2 S ) R^\mp \,, \\
    F_{n\bar{m}}^\pm &= \frac{2 \sqrt{2} \zeta^3}{\Delta \sin \theta} \bigg( \qty( \mathcal{L}^\dag_2 S ) \qty( \mathcal{D} R^\mp ) - \frac{(\zeta + \bar{\zeta})}{\zeta \bar{\zeta}} \qty( \mathcal{L}^\dag_2 S ) R^\mp + \frac{i a (\zeta - \bar{\zeta} ) \sin \theta}{\zeta \bar{\zeta}} S \qty( \mathcal{D} R^\mp ) \bigg) \,, \\
    F_{\bar{m}\bar{m}}^\pm &= - \frac{\zeta^3 }{\bar{\zeta} \sin^2\theta} S \qty( \mathcal{D}^2 R^\mp - \frac{2}{\zeta} \mathcal{D} R^\mp ) \,, \\
    G_n^\pm &= - \frac{\zeta^2 \bar{\zeta}}{\Delta} \qty(\mathcal{L}^\dag_1 \mathcal{L}^\dag_2 S - \frac{i a \sin\theta}{\zeta} \mathcal{L}^\dag_2 S ) \mathcal{D} R^\mp \,, \\
    G_{\bar{m}}^\pm &= \frac{\sqrt{2} \zeta^2 }{2\sin\theta} \mathcal{L}^\dag_2 S \qty( \mathcal{D}^2 R^\mp - \frac{1}{\zeta} \mathcal{D} R^\mp ) \,\
\end{align}
where
\begin{align}
    \mathcal{D} R^\mp &= \dv{R^\mp}{r} - i \frac{\mathcal{K}_r}{\Delta} R^\mp \,, \\
    \mathcal{L}^\dag_n S &= \dv{S}{\theta} + \frac{\mathcal{K}_\theta + n \cos\theta}{\sin\theta} S \,,
\end{align}
with $\mathcal{K}_r = (r^2+a^2)\omega - a m$, $\mathcal{K}_\theta = a \omega \sin^2 \theta - m$, and $R^\mp = R^\mp_{lm\omega}(r)$, $S = {}_{-2}S_{lm}^{a\omega}(\theta)$. For the $\omega$ derivatives of $F^\pm_{ab}$ we need the $\omega$ derivatives of $R^\mp$, $S$, $\mathcal{L}_2^\dag S$, $\mathcal{L}_1^\dag \mathcal{L}_2^\dag S$, $\mathcal{D}R^\mp$, and $\mathcal{D}^2 R^\mp$.

Expression of the geodesic source term \eqref{eq:Cpmlmnk_final} in terms of $F_{ab}$ is identical to rearranged expressions from \cite{Drasco:2006}. We can directly see that 
\begin{equation}
    F^\pm_{nn} v_n^2 = \dv{t}{\lambda} A_{nn0} R^\mp
\end{equation}
and after some simplifications that 
\begin{align}
    F^\pm_{n\bar{m}} v_n v_{\bar{m}} &= \dv{t}{\lambda} \qty( A_{nn0} R^\mp - A_{nn1} \dv{R^\mp}{r} ) \,, \\
    F^\pm_{\bar{m}\bar{m}} v_{\bar{m}}^2 &= \dv{t}{\lambda} \qty( A_{\bar{m}\bar{m}0} R^\mp - A_{\bar{m}\bar{m}1} \dv{R^\mp}{r} + A_{\bar{m}\bar{m}2} \dv[2]{R^\mp}{r} ) \,,
\end{align}
where $A_{abi}$ are the functions from Appendix B of \cite{Drasco:2006}.

\end{widetext}

\bibliography{main}

\end{document}